\documentclass{aastex}

\usepackage{spr-astr-addons}
\usepackage{url}
\usepackage{natbib}
\usepackage{here}

\RequirePackage{color}

\begin{document}

\title{Spectroscopic comparative study of the red giant binary system 
gamma Leonis A and B }
\shorttitle{Spectroscopic study of gamma Leonis A and B}
\shortauthors{Y. Takeda}

\author{Yoichi Takeda\altaffilmark{$\dagger$}}
\altaffiltext{$\dagger$}{11-2 Enomachi, Naka-ku, Hiroshima-shi 730-0851, Japan\\
E-mail: ytakeda@js2.so-net.ne.jp} 

\begin{abstract}
$\gamma$~Leo is a long-period visual binary system consisting of 
K0~{\sc iii} (A) and G7~{\sc iii} (B) giants, in which particular interest 
is attracted by the brighter A since the discovery of a planet around it. 
While detailed spectroscopic comparative study of both components would 
be worthwhile (e.g., for probing any impact of planet formation on chemical 
abundances), such a research seems to have been barely attempted as most 
available studies tend to be biased toward A. 
Given this situation, the physical properties of A and B along with their 
differences were investigated based on high-dispersion spectra in order 
to establish their stellar parameters, evolutionary status, and surface 
chemical compositions. The following results were obtained.
(1) The masses were derived as $\sim 1.7$~$M_{\odot}$ and $\sim 1.6$~$M_{\odot}$
for A and B, respectively, both of which are likely to be in the stage of 
red clump giants after He-ignition. The mass of the planet around A 
has also been revised as $m_{\rm p} \sin i_{\rm p} \simeq 10.7 M_{\rm Jupiter}$ 
(increased by $\sim 20\%$). 
(2) These are normal giants of subsolar metallicity ([Fe/H]~$\sim -0.4$) 
belonging to the thin-disk population.
(3) A as well as B show moderate C deficiency and N enrichment, which are
in compatible with the prediction from the standard stellar evolution theory.  
(4) The chemical abundances of 26 elements are practically the same
within $\la 0.1$~dex for both components, which implies that the surface 
chemistry is not appreciably affected by the existence of a planet in A.
\end{abstract}


\section{Introduction}

The star $\gamma$ Leo (Algieba)\footnote{This name, officially approved 
by the Working Group on Star Names of International Astronomical Union
in 2016, probably stemmed from the Arabic word ``{\it Al Jabbah}'' 
meaning ``forehead'', though this star is located rather 
in the ``mane'' of lion in most constellation charts.} 
is a well-known 2nd-magnitude star in the constellation Leo, 
which is known to be a visual binary consisting of two rather 
similar orange--yellowish stars (the brighter component $\gamma$~Leo~A =
$\gamma^{1}$~Leo = HD~89484 = HR~4057 with $V = 2.37$~mag and the fainter 
component $\gamma$~Leo~B = $\gamma^{2}$~Leo = HD~89485 = HR~4058 with 
$V = 3.64$~mag) separated by a few arcseconds after 
the discovery of Sir William Herschel in 1782. Since this system has a 
highly eccentric orbit (eccentricity as large as $e \sim 0.9$)  with very long period 
(several centuries), even half of its orbit has not been completed 
(neither periastron nor apoastron has been reached) over the past 240 years,
which means that its orbital elements still suffer considerable uncertainties.

Modern astronomical spectroscopy revealed that these component stars
are red giants classified as K0~{\sc iii} (A) and G7~{\sc iii} (B);
that is, the brighter former is somewhat redder than the fainter latter.
As these are apparently ordinary red giant stars like many others, this system 
was not of particular astrophysical interest for a long time, despite that 
it has been popular among amateur astronomers challenging to resolve it 
into two stars and to enjoy the delicate contrast of colors by using 
small home telescopes or binoculars.

However, a significant feature was recognized in this star about a decade ago, 
when Han et al. (2010) reported (based on the radial velocity method) 
the detection of a planetary companion with a mass of 
$m_{\rm p} \sin i_{\rm p} = 8.78 M_{\rm Jupiter}$ orbiting around $\gamma$~Leo~A 
with a period of 429 days. This is an important finding because it 
is presumably the first planet-hosting visual binary of two similar red giants 
which are separately observable.\footnote{According to the web site of 
``Extrasolar Planets Encyclopedia'', 173 planet-host binary systems are 
known as of 2023 February, among which only $\gamma$~Leo appears to meet 
this condition (cf. http://exoplanet.eu/planets\_binary/).  
}  

Generally speaking, a binary system in which only one component harbors 
a planet (while the other does not) is potentially an important testing 
bench for investigating the impact of planet formation on the host star, 
if the spectra of both stars are independently obtainable and their spectral 
types are not very different. That is, if any difference in the chemical 
abundances could be detected between the two, it would provide us with 
valuable information on the star--planet connection (e.g., accretion of 
proto-planetary materials), since they should have formed from gas 
with the same composition.

Although not a few such comparative studies of chemical abundances 
for the planet-host and non-planet-host components of visual binaries 
have been conducted so far, they are restricted to solar-type dwarfs
such as 16~Cyg A+B and HD~219542 A+B (see, e.g., Ryabchikova et al. 2022 
and the references therein). In this sense, it is worth determining 
the chemical abundances of many elements for $\gamma$~Leo A and B, 
in order to check whether any significant difference exists between 
these giant stars.
   
However, few such chemical abundance studies directed to both of 
$\gamma$~Leo A and B have been published. That is, most investigations 
on $\gamma$~Leo tend to be biased toward the brighter A,
while little attention has been paid to the fainter B. 
As a matter of fact, as summarized in Table~1, only two rather old studies 
are available that included $\gamma$~Leo~B (and also $\gamma$~Leo~A)
as one of the targets: 
(i) Lambert \& Ries's (1981) work on the CNO(+Fe) abundances of 
32 G--K giants, and (ii) McWilliam's (1990) abundance determinations 
of $\sim 10$ comparatively heavier elements (Si through Eu) 
for an extensive sample of 671 GK giants. 

\setcounter{table}{0}
\begin{table*}[h]
\caption{Stellar parameters of $\gamma$ Leo A and B reported in past publications.}
\scriptsize
\begin{center}
\begin{tabular}{ccccccl} 
\hline\hline
Authors & $T_{\rm eff}$ & $\log g$ & $v_{\rm t}$ & [Fe/H] & $M$ & Remark \\
\hline
\multicolumn{7}{c}{($\gamma$~Leo~A)} \\
Tomkin et al. (1975) & 4300 & 1.7  & 1.7  &  &  & $^{12}$C/$^{13}$C = 6.5 \\
Lambert \& Ries (1981) & 4650 & 2.39 & 2.0 & $-$0.35 &  &  \\
McWilliam (1990) & 4470 & 2.35 & 2.4 & $-$0.49 &  &  \\
Dyck et al. (1998) & 3949 &  &  &  &  & from interferometry-based angular diameter \\
$^{a}$Kovtyukh et al. (2006) & 4306 &  &  &  &  & assumed to be $\gamma$~Leo A (though labeled as HD~89485) \\
Cenarro et al. (2007) & 4470 & 2.12 &  & $-$0.38 &  &  \\
Massarotti et al. (2008) & 4365 & 2.3 &  & $-$0.49 &  &  \\
Han et al. (2010) & 4330 & 1.59 & 1.5 & $-$0.51 & 1.23  &  \\
Prugniel et al. (2011) & 4426 & 1.84 &  & $-$0.38 &  &  \\
Maldonado et al. (2013) & 4372 & 1.66 & 1.43 & $-$0.44 & 1.50  &  \\
Santos et al. (2013) & 4428 & 1.97 & 1.74 & $-$0.41 &  &  \\
Jofr\'{e} et al. (2015) & 4465 & 2.12 & 1.92 & $-$0.51 &  & $v_{\rm e}\sin i = 2.36$~km~s$^{-1}$  \\
Sousa et al. (2015) & 4395 & 1.66 & 1.67 & $-$0.47 & 1.46  &  \\
$^{a}$da Silva et al. (2015) & 4454 & 1.91 & 1.63 & $-$0.42 & $^{c}$(0.75)  & assumed to be $\gamma$~Leo A (though labeled as HD~89485) \\
Maldonado \& Villaver (2016) & 4376 & 1.67 & 1.43 & $-$0.44 & 1.59  &  \\
$^{b}$J\"{o}nsson et al. (2017) & 4341 & 1.77 & 1.56 & $-$0.48 &  & labeled as HIP~50583 $\rightarrow$ understood as $\gamma$~Leo A \\
$^{b}$Lomaeva et al. (2019) & 4341 & 1.77 & 1.56 & $-$0.45 &  & labeled as HIP~50583 $\rightarrow$ understood as $\gamma$~Leo A \\
Charbonnel et al. (2020) & 4274 &  &  & $-$0.54 & 1.77  & $A$(Li) = $-0.58$ \\
\hline
\multicolumn{7}{c}{($\gamma$~Leo~B)} \\
Lambert \& Ries (1981) & 5110 & 2.76 & 2.0 & $-$0.37 &  &  \\
McWilliam (1990) & 4980 & 2.98 & 2.5 & $-$0.52 &  &  \\
\hline
\end{tabular}
\end{center}
Summarized here are the effective temperature (in K), logarithmic surface gravity
in c.g.s unit (in dex), microturbulence (in km~s$^{-1}$), Fe abundance relative to the Sun, 
and mass (in $M_{\odot}$) taken from various previous studies. \\
$^{a}$Although the star is labeled as HD~89485 (which literally means $\gamma$~Leo B) in their table,
it is suspected to be a mistype of HD~89484 ($\gamma$~Leo A) as judged from the given $T_{\rm eff}$
value as low as $\la 4500$~K. Besides, it seems rather unnatural to do an analysis 
only for the fainter B without touching the brighter A. Therefore, it is tentatively assumed 
that these data should be understood as those of $\gamma$~Leo A.\\
$^{b}$Although the star is labeled simply as HIP~50583, this number in the Hipparcos catalogue
corresponds to $\gamma$~Leo A+B system as a whole (not the individual components). Therefore, 
it is assumed here that these data are those of the brighter $\gamma$~Leo A.\\
$^{c}$Presumably, some kind of error is involved in this mass value, which is too low. 
\end{table*}

Our group has so far published a series of papers focusing the abundances 
of various elements for a large sample of red giant stars: 
Takeda et al. (2008; hereinafter referred to as T08) [atmospheric parameters 
and abundances of many elements],
Takeda \& Tajitsu (2014; T14) [Be abundances],  
Takeda et al. (2015; T15) [C, O, and Na abundances],
Takeda et al. (2016; T16) [S and Zn abundances],
Takeda \& Tajitsu (2017; T17) [Li abundances], and
Takeda et al. (2019; T19) [$^{12}$C/$^{13}$C ratios and N abundances].
However, neither $\gamma$~Leo A nor B were included in our previous targets.
Therefore, this unsatisfactory situation motivated the author to newly carry 
out a detailed spectroscopic comparative study of both $\gamma$~Leo A and B 
in order to compare the chemical abundances for a number of elements 
(volatile as well as refractory elements) between these planet-host
and non-planet-host components, while making good use of our past experiences. 
This was the first motivation of this investigation.

Besides, by taking this opportunity, we intend to clarify the stellar 
parameters and the properties (e.g., activity level, kinematic information, 
etc.) of these two binary components, because they are not necessarily 
well established. Especially, although information of the stellar mass 
is important (which controls the stellar evolution and directly affects the 
mass evaluation for the orbiting planet), published results are rather diversified
($M$ of $\gamma$~Leo~A ranges from $\sim 1.2 M_{\odot}$ to $\sim 1.8 M_{\odot}$;
cf. Table~1). It should thus be worthwhile to determine the masses of both 
A and B as precisely as possible, such as attempted recently by Takeda (2022). 
for giants in the {\it Kepler} field.

In addition, we can also check the nature of internal mixing in $\gamma$~Leo  
based on the surface abundances of light elements (especially C, N, O, and Na). 
Our previous studies (T15 and T19) suggested that 
the surface abundance characteristics of mid-G to early-K giants 
(i.e., moderate deficiency in C, near-normal O, enrichment in N, 
mild overabundance in Na) are almost consistent with the results of 
recent theoretical simulations for red giants having experienced 
the first dredge-up (e.g., Lagarde et al. 2012).
Meanwhile, some previous work on $\gamma$~Leo~A done in 1990s reported 
that the surface abundances of the relevant light elements are anomalous
and in conflict with the standard theory. That is, nitrogen is nearly normal
([N/Fe]~$\sim -0.1$; Shavrina et al. 1996a) and carbon is even somewhat
overabundant ([C/Fe]~$\sim +0.2$; Shavrina et al. 1996b); this tendency
is apparently incompatible with the results corroborated in our past papers. 
Is $\gamma$~Leo~A a peculiar star in comparison with other red giant stars in general?  
To clarify this point is also counted as one of the tasks of this investigation.  

\section{Observational data}

The observational data (high-dispersion spectra) employed in this study were 
obtained in two observatories: Okayama Astrophysical Observatory (OAO) 
and Subaru Telescope (Subaru).
Actually, most of the analysis was done based on the former OAO spectra
covering the visible (and near IR) wavelength region,
while the latter Subaru spectra in the violet--UV region were subsidiarily 
used only for the specific purposes of Be abundance determination 
(from Be~{\sc ii} 3131) and measurement of Ca~{\sc ii} 3934 core emission.

\subsection{OAO observation}

The spectroscopic observations of $\gamma$~Leo A and B in the visible to
photographic IR wavelength region were done in 2010 May 3 (UT) by using  
HIDES (HIgh Dispersion Echelle Spectrograph) placed at the coud\'{e} 
focus of the 188 cm reflector at Okayama Astrophysical Observatory.
The exposure times were 300~s (A) and 1200~s (B). 
Equipped with three mosaicked 4K$\times$2K CCD detectors
at the camera focus, HIDES enabled us to obtain an echellogram covering 
$\sim$~5100--8800~\AA\ with a resolving power of $R \sim 67000$ 
(with the slit width of $200~\mu$m). 

\subsection{Subaru observation}

The Subaru observations of $\gamma$~Leo A and B  were carried out on 2010 
May 25 (UT) with HDS (High Dispersion Spectrograph) placed at the 
Nasmyth platform of the 8.2-m Subaru Telescope, by which high-dispersion 
spectra covering $\sim$~3000--4600~\AA\ could be obtained with two CCDs 
of 2K$\times$4K pixels in the standard Ub setting with 
the blue cross disperser. 
The spectrum resolving power was $R \simeq 60000$ with the slit width 
set at $0.''6$ (300 $\mu$m) and a binning of 2$\times$2 pixels. 
The integrated exposure times were 35~s (A) and 80~s (B), 
while the  star lights were considerably reduced with the help 
of a neutral density filter in order to avoid saturation. 

\subsection{Data reduction}

The reduction of the spectra (bias subtraction, flat-fielding, 
scattered-light subtraction, spectrum extraction, wavelength 
calibration, and continuum normalization) was performed by using 
the ``echelle'' package of the software IRAF\footnote{
IRAF is distributed by the National Optical Astronomy Observatories,
which is operated by the Association of Universities for Research
in Astronomy, Inc. under cooperative agreement with the National 
Science Foundation.} in a standard manner. 

\section{Stellar parameters}

\subsection{Atmospheric parameters}

The atmospheric parameters [effective temperature ($T_{\rm eff}$), 
surface gravity ($\log g$, where $g$ is in cm~s$^{-2}$), and 
microturbulence ($v_{\rm t}$) were spectroscopically determined 
in the same manner as in T08 (see Sect.~3.1 therein for the details) 
based on the equivalent widths ($W_{\lambda}$) of Fe~{\sc i} and Fe~{\sc ii} 
lines measured on the OAO spectra covering $\sim$~5100--8800~\AA. 
The resulting parameters for $\gamma$~Leo~A / $\gamma$~Leo~B are 
$4457(\pm 23)$ / $4969(\pm 15)$~K, $1.89(\pm 0.10)$ / $2.53(\pm 0.05)$~dex, 
and $1.44(\pm 0.10)$ / $1.39(\pm 0.08)$~km~s$^{-1}$,
where $\pm$ values in parentheses are internal statistical errors
(cf. Sect.~5.2 in Takeda et al. 2002).
The Fe abundances ($A_{\rm Fe}$)\footnote{
$A_{\rm X}$ is the logarithmic number abundance of element X, 
normalized with respect to H as $A_{\rm H} = 12.00$.}
 corresponding to the final solutions 
are plotted against $W_{\lambda}$ and $\chi_{\rm low}$ in Fig.~1,
where we can see that there is no systematic dependence as required.  
The detailed $W_{\lambda}$ and $A_{\rm Fe}$ data for each star are given 
in ``feabunds.dat'' of the supplementary material.
The mean Fe abundances ($\langle A_{\rm Fe} \rangle$) for A / B are 
$7.09(\pm 0.03)$ / $7.12(\pm 0.02)$, where $\pm$ values in parentheses 
are the mean errors ($\epsilon \equiv \sigma/\sqrt{N}$; $\sigma$ is the 
standard deviation and $N$ is the number of lines). The corresponding  
values of metallicity ([Fe/H])\footnote{
As usual, [X/H] is the differential abundance for element X of a star 
relative to the Sun; i.e., [X/H] $\equiv A_{\rm X\,*} - A_{\rm X\,\odot}$).
Likewise, the notation [X/Y] is defined as [X/Y] $\equiv$ [X/H] $-$ [Y/H].
Here, the relevant solar Fe abundance is $A_{\rm Fe\,\odot} = 7.50$.
} are $-0.41$ (A) and $-0.38$ (B).
The model atmosphere for each star to be used in this study was generated
by interpolating Kurucz's (1993) ATLAS9 model grid in terms of
$T_{\rm eff}$, $\log g$, and [Fe/H]. 

\setcounter{figure}{0}
\begin{figure}[H]
\begin{minipage}{80mm}
\begin{center}
\includegraphics[width=8.0cm]{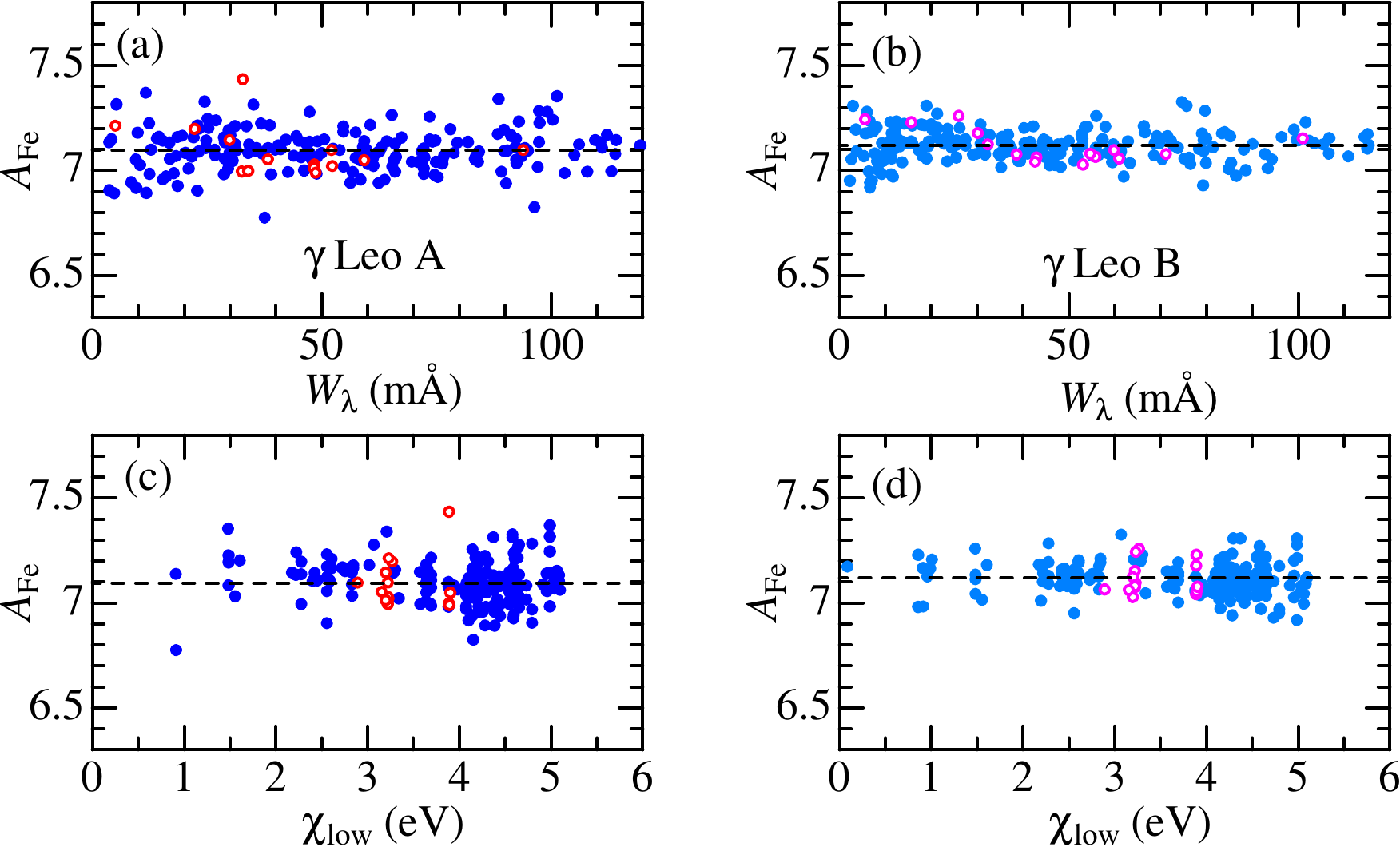}
\caption{
Upper panels (a) and (b): $A_{\rm Fe}$ (Fe abundance) vs. $W_{\lambda}$ 
(equivalent width) relations. Lower panels (c) and (d):  $A_{\rm Fe}$ vs. 
$\chi_{\rm low}$ (lower excitation potential) relations.  
These Fe abundances correspond to the finally established atmospheric parameters 
of $T_{\rm eff}$, $\log g$, and $v_{\rm t}$ for $\gamma$~Leo~A (left panels)
and $\gamma$~Leo~B (right panels). 
The filled and open symbols correspond to Fe~{\sc i} and Fe~{\sc ii} 
lines, respectively. The mean abundance ($\langle A_{\rm Fe}\rangle$) 
is indicated by the horizontal dashed line. 
}
\label{fig1}
\end{center}
\end{minipage}
\end{figure}

\subsection{Mass and age}

The absolute magnitude ($M_{V}$) or luminosity ($L$)
are determinable from the apparent magnitude ($V$) and 
the parallax ($\pi$) with appropriate corrections, Then, since 
the position on the theoretical HR diagram is established in 
combination with $T_{\rm eff}$, the mass ($M$) as well as 
stellar age ($age$) can be evaluated with the help of theoretical 
evolutionary tracks.  

For this purpose, the open software tool PARAM version 1.3\footnote{
http://stev.oapd.inaf.it/cgi-bin/param\_1.3/} 
(da Silva et al. 2006) was employed as done by Takeda (2022), 
which requires $T_{\rm eff}$, [Fe/H], $V_{0} (\equiv V - A_{V})$, 
and $\pi$ (along with their errors) as input parameters (see Table~2). 
The uncertainty in $V - A_{V}$ was assumed to be 0.05~mag (that of $A_{V}$).
The output results of $M$, $age$, $R$ (radius), and $\log g_{MR}$ 
(gravity from $M$ and $R$) are also summarized in Table~2.

For the sake of confirming these solutions, the positions of $\gamma$~Leo 
A and B on the $\log T_{\rm eff}$ vs. $\log L$ diagram are compared
with the theoretical evolutionary tracks in Fig.~2, and the
PARAM results of $age$ vs. $M$ relation are illustrated in Fig.~3.

Regarding the evolutionary status of these two stars, although it is 
difficult to discriminate based on these figures whether they are 
in the H-burning phase (ascending the red giant branch) or in the 
post-He-ignition phase (red clump giants), the latter would be more likely 
for both, because of the sign of advanced first dredge-up (such as 
the low $^{12}$C/$^{13}$C ratio around $\sim 10$; cf. Sect.~5.1.2).

\setcounter{figure}{1}
\begin{figure}[H]
\begin{minipage}{80mm}
\begin{center}
\includegraphics[width=6.0cm]{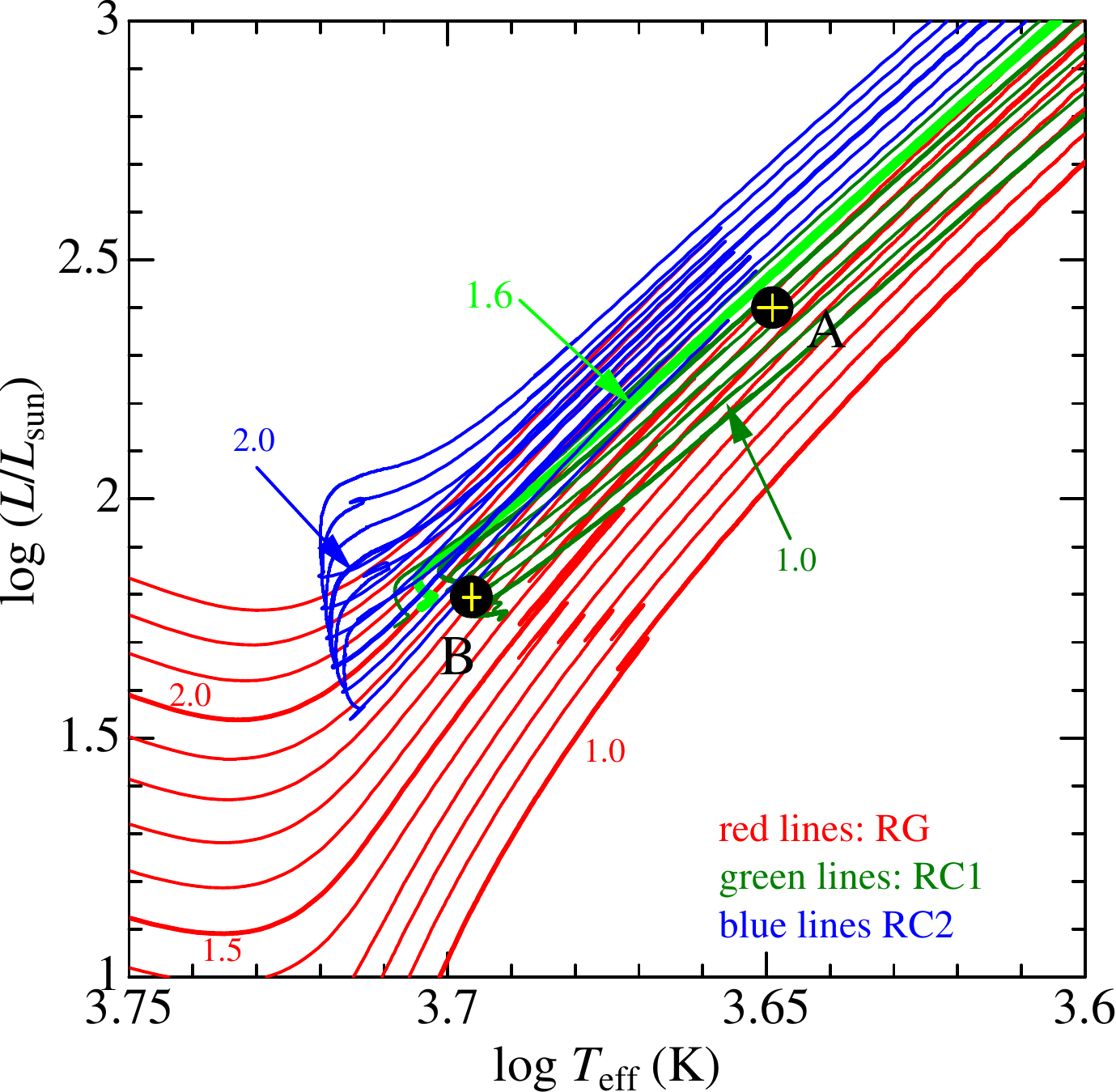}
\caption{
Positions of $\gamma$~Leo A and B (plotted by large filled circles 
with error bars indicated in yellow crosses) 
on the $\log T_{\rm eff}$--$\log L$ diagram, where Bressan et al.'s (2012) PARSEC 
tracks computed for $z = 0.006$ (moderately metal-deficient by $-0.37$~dex 
lower than the solar metallicity) and various $M$ values 
(1.0, 1.1, 1.2, 1.3, 1.4, 1.5, 1.6, 1.7, 1.8, 1.9, 2.0, 2.1, 2.2, 
and 2.3~$M_{\odot}$;  thick lines for the cases of integer masses, 
otherwise thin lines) are overplotted for comparison. 
These theoretical tracks are depicted in different colors corresponding to 
their evolutionary stages: Red $\cdots$ shell-H-burning phase before He ignition 
(Red Giant phase or RG). 
Green $\cdots$ $M \le 1.8$~$M_{\odot}$ stars at the He-burning phase 
(1st Red Clump giant phase or RC1). 
Blue $\cdots$ $M > 1.8$~$M_{\odot}$ stars at the He-burning phase 
(2nd Red Clump phase or RC2). 
}
\label{fig2}
\end{center}
\end{minipage}
\end{figure}

\setcounter{figure}{2}
\begin{figure}[H]
\begin{minipage}{80mm}
\begin{center}
\includegraphics[width=5.0cm]{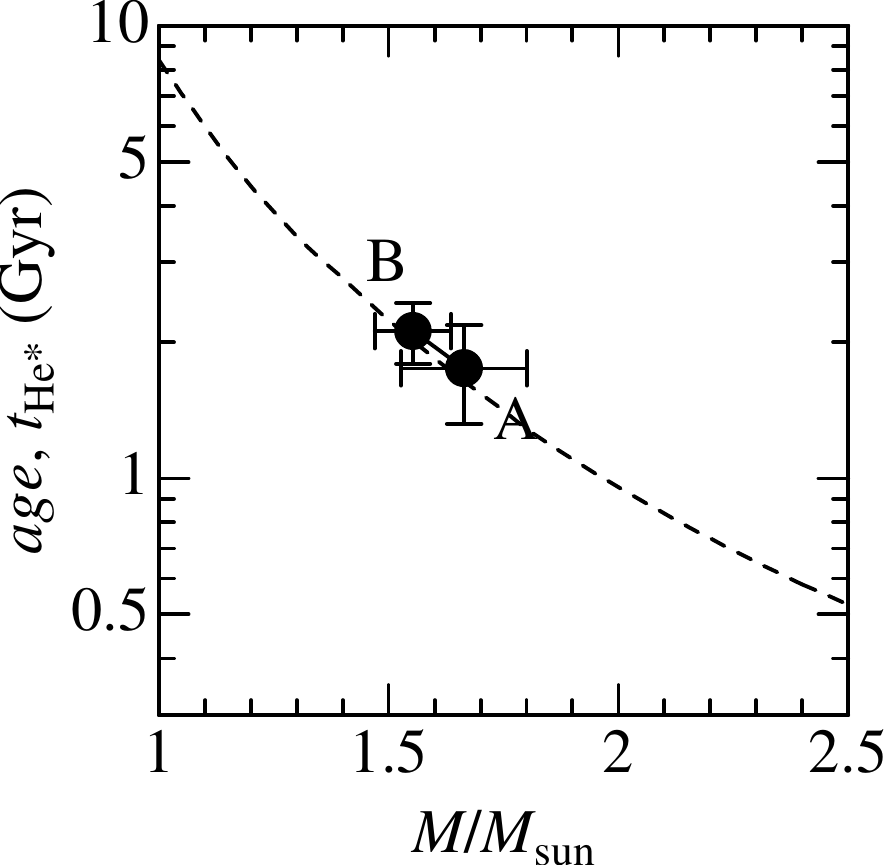}
\caption{
PARAM results of $age$ ($t$) and mass ($M$) for $\gamma$~Leo A and B 
plotted on the $age$--$M$ diagram, where $t_{\rm He*}$ (time of 
He-ignition) vs. $M$ relation (taken from PARSEC tracks) is 
also shown by the dashed line.
}
\label{fig3}
\end{center}
\end{minipage}
\end{figure}

\setcounter{figure}{3}
\begin{figure}[H]
\begin{minipage}{80mm}
\begin{center}
\includegraphics[width=5.0cm]{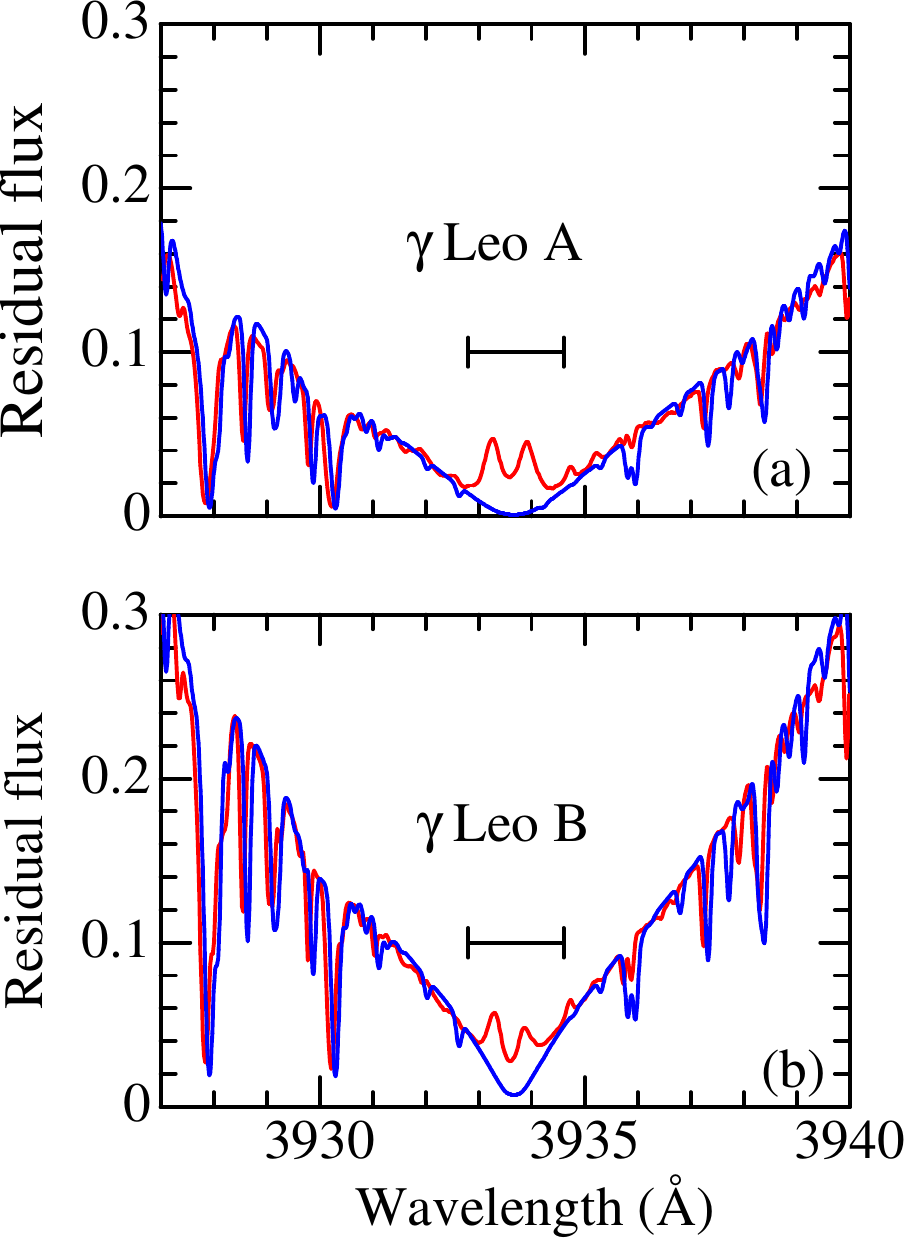}
\caption{
Observed residual flux spectra of Ca~{\sc ii} K line at 3933.663~\AA\ are 
depicted by red lines for (a) $\gamma$~Leo A and (b) $\gamma$~Leo B. 
In addition, theoretical spectra ($r_{\lambda}^{\rm th}$) calculated from 
the model atmospheres with the assumption of LTE are also overplotted by blue lines.
The core region (3932.8--3934.6~\AA), where integration was done 
for evaluating the emission strength, is indicated by a horizontal bar. 
The wavelength scale is adjusted to the laboratory frame by correcting 
the radial velocity shift. 
}
\label{fig4}
\end{center}
\end{minipage}
\end{figure}

\subsection{Stellar activity and rotation}

In order to estimate the chromospheric activity, the activity index
$\log R'_{\rm Kp}$ was evaluated from the Subaru spectra
by following the procedure described in Sect.~3 of Takeda et al. (2012)
where $R'_{\rm Kp}$ is the ratio of the chromospheric core emission 
flux of the Ca~{\sc ii} resonance line at 3933.663~\AA\ (after subtraction 
of the photospheric flux computed from the classical model atmosphere) 
to the total bolometric flux. Fig.~4 displays the appearance of the spectra 
in the relevant core region of Ca~{\sc ii} 3934 line. The resulting
$\log R'_{\rm Kp}$ values are $-5.11$ (A) and $-5.07$ (B). 

The projected rotational velocity ($v_{\rm e}\sin i$), which is closely 
related to stellar activity, was determined from the spectrum fitting
analysis in the 6080--6089~\AA\ region (cf. Fig.~5a and the top row of Table~3), 
as detailed in Sect.~4.2 of T08. The $v_{\rm e} \sin i$ results for
$\gamma$~Leo A and B are 1.41 and 1.62~km~s$^{-1}$, respectively.

Comparing these values of $\log R'_{\rm Kp}$ and $v_{\rm e}\sin i$ with the 
trends shown in Fig.~6c--e of T17, we can see that both $\gamma$~Leo A and B 
are typical red giant stars of low activity for their low rotational velocities.

\subsection{Kinematic parameters}

The velocity components of a star and the properties of its orbital motion 
in the Galaxy are important to understand the stellar population.
These kinematic parameters for the $\gamma$~Leo system (A and B do not need 
to be distinguished here) were obtained by following the procedure described 
in Sect.~2.2 of Takeda (2007), and the results are summarized in the last 
section of Table~2.

Plotting the resulting values of $V_{\rm LSR}$ (+11.7~km~s$^{-1}$) and 
$z_{\rm max}$ (0.063~kpc) on the $z_{\rm max}$ vs. $V_{\rm LSR}$ diagram 
(cf. Fig.~5a of T17), we can state that $\gamma$~Leo belongs to the ordinary 
thin-disk population. This conclusion was also confirmed by applying 
the $U_{\rm LSR}$, $V_{\rm LSR}$ and $W_{\rm LSR}$ values to 
Bensby et al.'s (2005) kinematical criteria (cf. Appendix~A therein),
which yields $TD/D = 0.05$ satisfying the criterion of $0 < TD/D < 1$
for thin-disk population.

\setcounter{table}{1}
\begin{table*}[h]
\caption{Fundamental parameters of $\gamma$ Leo A and B adopted/derived in this study.}
\scriptsize
\begin{center}
\begin{tabular}{ccccl} 
\hline\hline
Quantity & (Unit) & $\gamma$~Leo~A & $\gamma$~Leo~B & Explanation \\
\hline
 \multicolumn{4}{c}{(atmospheric parameters determined from Fe~{\sc i}/Fe~{\sc ii} lines)} & \\ 
$T_{\rm eff}$ & (K) & 4457 ($\pm 23$) & 4969 ($\pm 15$) & effective temperature \\
$\log g$ & (dex) & 1.89 ($\pm 0.10$) & 2.53 ($\pm 0.05$)& surface gravity (in c.g.s.) \\
$v_{\rm t}$ & (km~s$^{-1}$) & 1.44 ($\pm 0.10$)& 1.39 ($\pm 0.08$)& microturbulence \\
$[{\rm Fe/H}]$ & (dex) & $-0.41$ ($\pm 0.03$)& $-0.38$ ($\pm 0.02$)& metallicity \\
\hline
 \multicolumn{4}{c}{(Gaussian macroturbulence and projected rotational velocity)} & \\ 
$v_{\rm mt}$      & (km~s$^{-1}$) & 3.03 & 2.60 & based on 6080--6089~\AA\ fitting (cf. Eq.~1 in T08) \\
$v_{\rm e}\sin i$ & (km~s$^{-1}$) & 1.41 & 1.62 & based on 6080--6089~\AA\ fitting (cf. Eq.~2 in T08) \\
\hline
 \multicolumn{4}{c}{(activity index)} & \\ 
$\log R'_{\rm Kp}$ & (dex) & $-5.11$ & $-5.07$ & see Eq.~2 in Takeda et al. (2012) for its definition\\
\hline
 \multicolumn{4}{c}{(photometric parameters)} & \\ 
$V$ & (mag) & 2.37 & 3.64 & apparent magnitude in $V$-band (from SIMBAD)\\
$\pi$ & (m.a.s.) & 25.07 ($\pm 0.52$) & 25.07 ($\pm 0.52$)& parallax from revised Hipparcos catalogue\\
$A_{V}$ & (mag) & 0.10 ($\pm 0.05$) & 0.10 ($\pm 0.05$)& interstellar extinction from EXTINCT (Hakkila et al. 1997)\\
B.C. & (mag) & $-0.52$ & $-0.27$ & bolometric correction according to Alonso et al. (1999)\\
$\log(L/L_{\odot})$ & (dex) & 2.40 & 1.80 & bolometric luminosity\\
\hline
  \multicolumn{4}{c}{(Parameters determined from evolutionary tracks)} & \\ 
$M$ & ($M_{\odot}$) & 1.66 ($\pm 0.14$) & 1.55 ($\pm 0.08$)& mass (by PARAM)\\
$age$ & (Gyr) & 1.75 ($\pm 0.43$) & 2.12 ($\pm 0.33$)&  age (by PARAM)\\
$R$  & ($R_{\odot}$) & 26.08 ($\pm 0.79$)& 10.55 ($\pm 0.29$)& radius (by PARAM)\\  
$\log g_{MR}$ & (dex) & 1.80 ($\pm 0.04$) & 2.56 ($\pm 0.04$)& surface gravity from $M$ and $R$ (by PARAM)\\
\hline
  \multicolumn{4}{c}{(kinematic parameters)} & \\ 
$v_{\rm r}^{\rm hel}$ & (km~s$^{-1}$) & \multicolumn{2}{c}{$-36.24$} & heliocentric radial velocity (from SIMBAD) \\
$\mu_{\alpha}$ & (m.a.s.~yr$^{-1}$) & \multicolumn{2}{c}{$+304.30$} & proper motion in $\alpha$ direction (from SIMBAD) \\
$\mu_{\delta}$ & (m.a.s.~yr$^{-1}$) & \multicolumn{2}{c}{$-154.28$} & proper motion in $\delta$ direction (from SIMBAD) \\
$U_{\rm LSR}$ & (km~s$^{-1}$) & \multicolumn{2}{c}{84.2} & radial component of space velocity at LSR (Local Standard of Rest)\\ 
$V_{\rm LSR}$ & (km~s$^{-1}$) & \multicolumn{2}{c}{11.7} & tangential component of space velocity at LSR\\
$W_{\rm LSR}$ & (km~s$^{-1}$) & \multicolumn{2}{c}{3.6}  & vertical component of space velocity at LSR\\ 
$|v|_{\rm LSR}$ & (km~s$^{-1}$) & \multicolumn{2}{c}{85.1} & $\sqrt{U_{\rm LSR}^{2}+V_{\rm LSR}^{2}+W_{\rm LSR}^{2}}$\\ 
$R_{\rm m}$ & (kpc) & \multicolumn{2}{c}{8.938} & mean galactocentric radius\\ 
$e$ & $\cdots$ & \multicolumn{2}{c}{0.262} & orbital ellipticity\\ 
$z_{\rm max}$ & (kpc) & \multicolumn{2}{c}{0.063} & maximum separation from the galactic plane\\
\hline
\end{tabular}
\end{center}
\end{table*}

\section{Determination of elemental abundances}

\subsection{Basic policy}

In this study, larger weight is put to comparatively lighter elements 
(rather than heavier ones), because important volatile species or 
mixing-sensitive elements are included in this group.
Therefore,  abundances of such lighter elements are derived based on 
the spectrum-fitting method by taking the non-LTE effect into account,
while abundance determinations for heavier elements are done by the 
conventional manner using equivalent widths with the assumption of LTE. 

We exclusively focus on the relative abundances of 
$\gamma$~Leo A and B in comparison with the Sun\footnote{
Strictly speaking, our Sun may not necessarily be adequate as the reference 
standard, because its surface composition tends to shows a marginally 
atypical signature. Mel\'{e}ndez et al.~(2009)
reported in their high-precision differential study of nearby solar twins 
in comparison with the Sun that the solar abundances of 
refractory elements (such as Fe group) are slightly deficient relative to the 
volatile ones  (such as CNO), which might be associated with the formation 
mechanism of our solar system (especially rocky terrestrial planets).
However, we do not need to care about this problem in this study, since 
the magnitude of this effect (on the order of several hundredths dex) 
is not significant as compared to the typical precision of abundance 
determination ($\la 0.1$~dex).
} 
([X/H]$_{\rm A}$ or ([X/H]$_{\rm B}$) along with their mutual differences 
($\delta$[X/H]$_{{\rm A}-{\rm B}}$). 
Since they are derived by applying the differential analysis under the same
condition (spectrum-fitting done in the same manner or differential line-by-line 
analysis based on equivalent widths) to the three spectra of A, B, and the Sun. 
uncertainties in atomic line parameters (especially those of oscillator strengths) 
are cancelled out and thus irrelevant.

\subsection{Spectrum fitting analysis}

Such as was conducted in T17, the abundances of 7 elements (Li, Be, C, O, 
Na, S, and Zn; either important lighter elements or volatile species) 
for $\gamma$~Leo A and B (as well as for the Sun) were determined by 
applying the spectrum-fitting technique, followed by a non-LTE analysis 
(except for several cases where LTE was assumed) based on the equivalent 
widths inversely derived from the best-fit abundance solutions 
(see Sect.~7--9 in T17 for more explanations of the procedures).
Likewise, CN abundances (usable to obtain the abundances of N) 
and $^{12}$C/$^{13}$C ratios were derived by the synthetic 
fitting method as done in T19.
 
Specific information (wavelength range, varied abundances, reference sources 
of the line data) regarding these spectrum fitting analyses is summarized 
in Table~3, and the atomic data for the representative key lines in each of
the wavelength regions are presented in Table~4.  
Regarding the spectra for the Sun (to be used for deriving the reference 
solar abundances), the OAO spectra of Moon in the visible--near IR region 
(as in T15) and the Subaru spectra of Vesta in the UV--violet region 
(as in T17) were adopted.

How the theoretical and observed spectra match each other is displayed 
for each region in Fig.~5, and the results of the analysis (equivalent widths, 
non-LTE correction, elemental abundances, and differential abundances 
relative to the Sun) are summarized in Table~5.

\setcounter{table}{2}
\setlength{\tabcolsep}{3pt}
\begin{table*}[h]
\scriptsize
\caption{Spectrum fitting analysis done in this study.}
\begin{center}
\begin{tabular}{ccccc}\hline\hline
Purpose & Fitting range (\AA) & Abundances varied$^{*}$ & Atomic data source & Figure$^{\dagger}$ \\
\hline
$v_{\rm e}\sin i$ determination & 6080--6089  & Si, Ti, V, Fe, Co, Ni &KB95 (cf. T08)  & a \\
Li abundance from Li~{\sc i} 6708  & 6707--6708.5 &  Li, Fe, V & SLN98+VALD (cf. T17) & b \\
Be abundance from Be~{\sc ii} 3131 & 3130.45--3131.2 &  OH, Be, Ti, (Fe)$_{\rm fix}$ & P97 (cf. T14) & c \\
C abundance from C~{\sc i} 5052   & 5051.3--5052.4 & C, Cr, Fe, Ni & KB95m1 & d \\
C abundance from C~{\sc i} 5380   & 5378.5--5382 & C, Ti, Fe, Co & KB95 (cf. T15) & e \\
C abundance from C~{\sc i} 8335   & 8333.5--8336 & C, Ti & KB95 & f \\
C abundance from [C~{\sc i}] 8727   & 8726--8730 & C, Si, Fe & KB95 (cf. T15) & g \\
CN and $^{12}$C/$^{13}$C from CN lines  &  8001--8006 & CN, Fe, $^{12}$C/$^{13}$C & CCSM+SJIS (cf. T19) & h \\ 
O abundance from O~{\sc i} 7771--5   & 7770--7777 & O, Fe, Nd, (CN)$_{\rm fix}$ & TKS98+KB95 (cf. T15) & i \\
Na abundance from Na~{\sc i} 6161  & 6157--6164 & Na, Ca, Fe, Ni & KB95 (cf. T15) & j \\
S abundance from S~{\sc i} 6757   & 6756--6758.1 & S, Fe, Co & KB95 (cf. T16) & k \\
Zn abundance from Zn~{\sc i} 6362  & 6361--6365 & O, Cr, Fe, Ni, Zn & KB95 (cf. T16) & l \\
\hline
\end{tabular}
\end{center}
$^{*}$ The abundances of other elements than these were fixed by assuming [X/H] = [Fe/H] in the fitting.\\ 
$^{\dagger}$ Corresponding panel of Fig.~5.\\
CCSM --- Carlberg et al. (2012),
KB95 --- Kurucz \& Bell (1995), 
KB95m1 --- Kurucz \& Bell (1995) without the Fe~{\sc i} 5051.276 line, 
SLN98 --- Smith, Lambert, \& Nissen (1998),
VALD --- VALD database (Ryabchikova et al. 2015),
P97 --- Primas et al. (1997),
SJIS --- Sablowski et al. (2019), and
TKS98 --- Takeda, Kawanomoto, \& Sadakane (1998). 
\end{table*}

\setcounter{table}{3}
\setlength{\tabcolsep}{3pt}
\begin{table}[h]
\scriptsize
\caption{Line data adopted for evaluating the equivalent widths.}
\begin{center}
\begin{tabular}{cccccl}\hline\hline
Line & $W$ & $\lambda$ & $\chi_{\rm low}$ & $\log gf$ & Remark\\
     &     & (\AA) & (eV) & (dex) & \\  
\hline
Li~{\sc i} 6708& $W_{6708}$ & 6707.756 & 0.00 & $-0.427$ & $^{7}$Li \\
 & & 6707.768 & 0.00 & $-0.206$ &  $^{7}$Li\\
 & & 6707.907 & 0.00 & $-0.932$ &  $^{7}$Li\\
 & & 6707.908 & 0.00 & $-1.161$ &  $^{7}$Li\\
 & & 6707.919 & 0.00 & $-0.712$ &  $^{7}$Li\\
 & & 6707.920 & 0.00 & $-0.932$ &  $^{7}$Li\\
\hline
Be~{\sc ii} 3131 & $W_{3131}$ & 3131.066 & 0.000 & $-0.468$ & $^{9}$Be \\
\hline
C~{\sc i} 5052 & $W_{5052}$  & 5052.167 & 7.685 & $-1.648$ & \\
\hline
C~{\sc i} 5380 & $W_{5380}$  & 5380.337 & 7.685 & $-1.842$ & \\
\hline
C~{\sc i} 8335 & $W_{8335}$  & 8335.147 & 7.685 & $-0.437$ & \\
\hline
[C~{\sc i}] 8727 & $W_{8727}$  & 8727.126 & 1.264 & $-8.210$ & \\
\hline
O~{\sc i} 7774 & $W_{7774}$ & 7774.166 & 9.146 & +0.174 & \\
\hline
Na~{\sc i} 6161 & $W_{6161}$ & 6160.747 & 2.104 & $-1.260$ & \\
\hline
S~{\sc i} 6757 & $W_{6757}$ & 6756.851 & 7.870 & $-1.760$ & 3 components\\
 & & 6757.007 & 7.870 & $-0.900$ & \\
 & & 6757.171 & 7.870 & $-0.310$ & \\
\hline
Zn~{\sc i} & $W_{6362}$ & 6362.338 & 5.796 & +0.150 & \\ 
\hline
\end{tabular}
\end{center}
In columns 3--5 are presented the atomic line data of $\lambda$ (air wavelength), 
$\chi_{\rm low}$ (lower excitation potential) and $\log gf$ (logarithm of
statistical weight times oscillator strength), respectively.
See Table~3 for the reference sources of these data. 
\end{table}

\setcounter{table}{4}
\setlength{\tabcolsep}{3pt}
\begin{table*}[h]
\scriptsize
\caption{Abundance results derived from spectrum fitting.}
\begin{center}
\begin{tabular}{cccccccl}\hline\hline
 Line  & Star & $W_{\lambda}$ & $\Delta$ & $A_{\rm X}$ & N/L & [X/H] & Remark \\
 (1)   & (2)  &   (3)         &  (4)     & (5) &    (6)      &  (7)  & (8) \\
\hline
Li~{\sc i}~6708 & A & (1.6)   & $\cdots$ & ($-$0.81) & L & $\cdots$ & Detection limit (upper limit)\\
                & B & (0.9)   & $\cdots$ & ($-$0.36) & L & $\cdots$ & Detection limit (upper limit)\\
\hline                                                      
Be~{\sc ii}~3131& A & 47.5  & $\cdots$ & $-$0.234 & L & $\cdots$ & Fe fixed, less reliable\\
                & B & 61.0  & $\cdots$ & 0.078    & L & $\cdots$ & Fe fixed, less reliable\\
\hline                                                      
C~{\sc i}~5052  & A &  8.7  & $-$0.025 &  8.133  & N & $-$0.531 &  \\
                & B & 12.3  & $-$0.023 &  8.080  & N & $-$0.584 &  \\
                & S & 27.4  & $-$0.009 &  8.664  & N &          &  \\
\hline                                                      
C~{\sc i}~5380  & A &  9.9  & $-$0.028 &  8.458  & N & $-$0.214 &  Not used (maybe blended) \\
                & B & 10.3  & $-$0.024 &  8.207  & N & $-$0.465 &  Not used\\
                & S & 19.9  & $-$0.008 &  8.672  & N &          &  \\
\hline                                                      
C~{\sc i}~8335  & A & 21.3  & $-$0.071 &  7.769  & N & $-$0.770 &  \\
                & B & 43.7  & $-$0.102 &  7.814  & N & $-$0.725 &  \\
                & S & 105.4 & $-$0.068 &  8.539  & N &          &  \\
\hline                                                      
[C~{\sc i}]~8727& A & 6.6   & $\cdots$ &  7.923  & L & $-$0.564 &  \\
                & B & 7.3   & $\cdots$ &  7.875  & L & $-$0.612 &  \\
                & S & 4.9   & $\cdots$ &  8.487  & L &          &  \\
\hline                                                      
O~{\sc i}~7774  & A & 17.8  & $-$0.091 &  8.540  & N & $-$0.333 &  \\
                & B & 36.6  & $-$0.143 &  8.541  & N & $-$0.332 &  \\
                & S & 64.0  & $-$0.103 &  8.873  & N &          &  \\
\hline                                                      
Na~{\sc i}~6161 & A & 100.0 & $-$0.098 &  6.063  & N & $-$0.243 &  \\
                & B & 63.9  & $-$0.079 &  5.934  & N & $-$0.372 &  \\
                & S & 59.0  & $-$0.058 &  6.306  & N &          &  \\
\hline                                                      
S~{\sc i}~6757  & A & 6.4   & $-$0.022 &  6.842  & N & $-$0.355 &  \\
                & B & 9.4   & $-$0.021 &  6.747  & N & $-$0.450 &  \\
                & S & 19.9  & $-$0.005 &  7.197  & N &          &  \\
\hline                                                      
Zn~{\sc i}~6363 & A & 20.4  & $-$0.047 &  4.277  & N & $-$0.218 &  \\
                & B & 20.6  & $-$0.020 &  4.186  & N & $-$0.309 &  \\
                & S & 19.4  & $-$0.005 &  4.495  & N &          &  \\
\hline
\end{tabular}
\end{center}
(1) Line designation (same as in Table~4). (2) Key to the relevant star: A~$\cdots \gamma$~Leo~A, 
B~$\cdots \gamma$~Leo~B, and S~$\cdots$~Sun. (3) Equivalent width (in m\AA) inversely derived
from the abundance solution of spectrum fitting along with the atomic data in Table~4.
(4) Non-LTE correction ($\equiv A^{\rm NLTE} - A^{\rm LTE}$).
(5) Final abundance of element X (in the usual normalization of $A_{\rm H} = 12$). 
(6) L $\cdots$ LTE abundance. N $\cdots$ non-LTE abundance. 
(7) Differential abundance relative to the Sun; i.e., 
[X/H]$_{\rm A}$~$\equiv A_{\rm X}$(A)$-$$A_{\rm X}$(S) for A, 
and [X/H]$_{\rm B}$~$\equiv A_{\rm X}$(B)$-$$A_{\rm X}$(S) for B. 
(8) Additional remark.
\end{table*}

\subsection{Derivation of C and N abundances}

The role played by C is especially significant because it directly 
affects the abundance of N determinable from CN.
In this study, C abundanes were derived for 4 lines (C~{\sc i} 5052, 
C~{\sc i} 5380, C~{\sc i} 8335, and [C~{\sc i}] 8727).
An inspection of the abundance difference between A and B 
($\delta \equiv A^{\rm A} - A^{\rm B}$) revealed that only that 
for C~{\sc i} 5380 is appreciably large by $\delta \sim +0.25$~dex 
while those for the other three lines are as small as $|\delta| \la 0.05$~dex
(cf. Table~5). Since this suggests that the C~{\sc i} 5380 line is 
likely to be contaminated by blending of some other lines in cooler A 
(but not for the hotter B), this line was discarded in calculating 
the mean C abundances, which were derived from the other three lines
as [C/H] = $-0.62$ (A) and $-0.64$ (B); or [C/Fe] 
($\equiv [{\rm C/H}] - [{\rm Fe/H}]$) = $-0.22$ (A) and $-0.26$ (B).
Then, the N abundances are evaluated by combining [C/Fe] and 
$\phi_{\rm CN}$ (scale factor of CN) as described in Sect.~3.2
in T19. Such derived N abundances are summarized in Table~6.

\setcounter{table}{5}
\setlength{\tabcolsep}{3pt}
\begin{table*}[h]
\scriptsize
\caption{Results from the fitting analysis of CN molecular lines.}
\begin{center}
\begin{tabular}{cccccccc}\hline\hline
Star & [Fe/H] & $^{12}$C/$^{13}$C & $\log \phi_{\rm CN}$ & $[\phi_{\rm CN}]$ &
[C/Fe] & [N/Fe] & [N/H] \\
\hline
$\gamma$ Leo A & $-0.41$ & 8.6 & $-0.110$ & 0.044 & $-0.22$ & 0.26 & $-0.14$ \\
$\gamma$ Leo B & $-0.38$ &10.8 & $-0.097$ & 0.057 & $-0.26$ & 0.31 & $-0.07$ \\
\hline
\end{tabular}
\end{center}
$\phi_{\rm CN}$ is the depth-independent scale factor, by which the number population 
of CN molecules (computed from a model atmosphere with the metallicity-scaled abundances) 
is to be multiplied to reproduce the observed CN line strengths.
$[\phi_{\rm CN}] \equiv \log \phi_{\rm CN,*} - \log \phi_{\rm CN.\odot}$,
where $\log \phi_{\rm CN,\odot} = -0.154$ (cf. Sect.~3.2 in T19).
N abundances are derived by the relation $[{\rm N/Fe}] = [\phi_{\rm CN}] - [{\rm C/Fe}]$
(cf. Eq.~1 in T19).
$^{12}$C/$^{13}$C is the ratio of carbon isotopes.
\end{table*}

\setcounter{figure}{4}
\begin{figure}[H]
\begin{minipage}{80mm}
\begin{center}
\includegraphics[width=8.0cm]{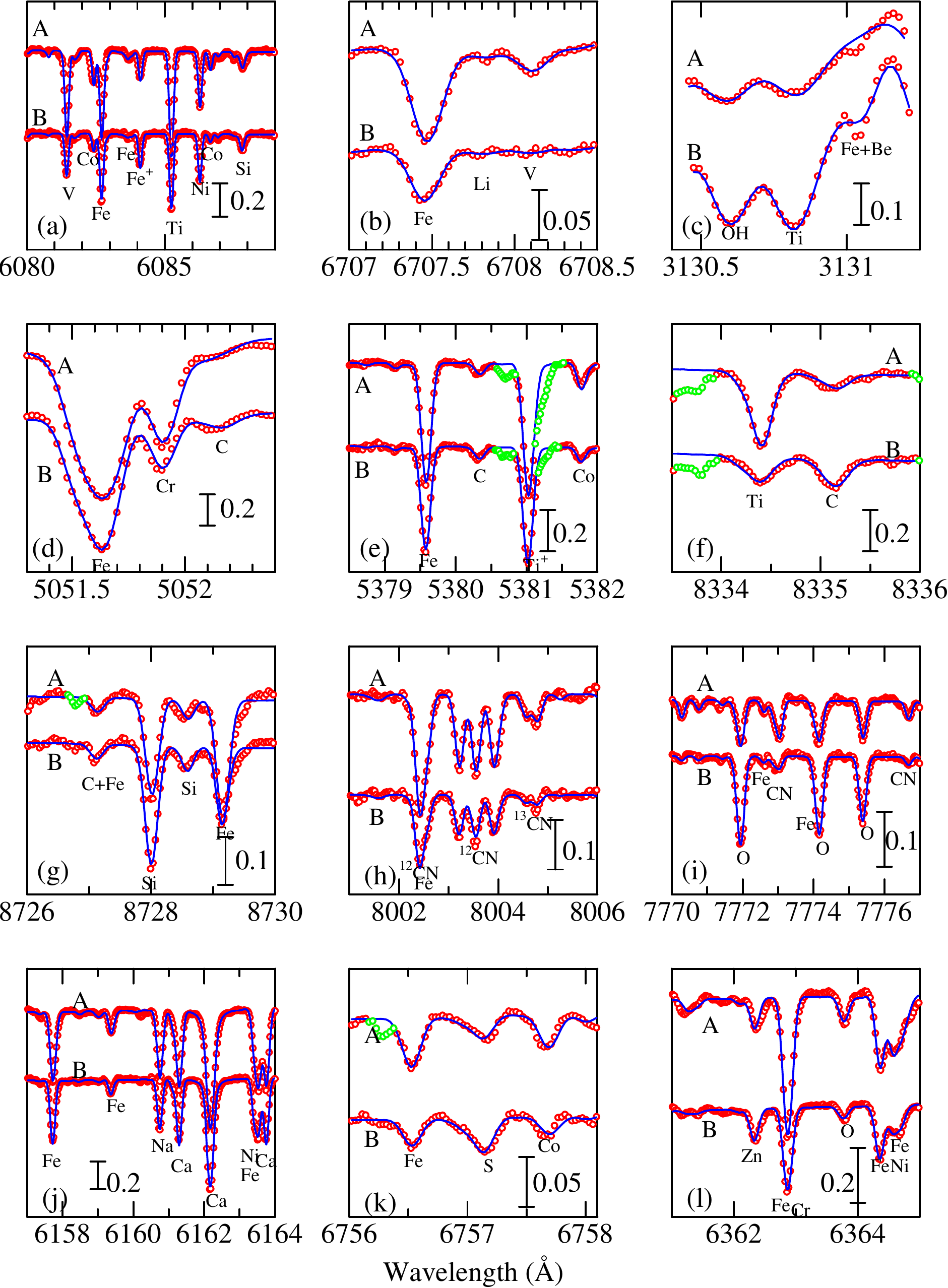}
\caption{
Synthetic spectrum fittings carried out (mainly) for determining 
the abundances of light elements.(cf. Table~3). 
The observed spectra are plotted in red symbols (where the masked regions 
discarded in judging the goodness of fit are colored in light-green)
while the best-fit theoretical spectra are shown in blue solid lines. 
The wavelength scale is adjusted to the laboratory system.  
(a) 6080--6089~\AA\ region,
(b) 6707--6708.5~\AA\ region,
(c) 3130.45--3131.2~\AA\ region, 
(d) 5051.3--5052.4~\AA\ region,
(e) 5378.5--5382~\AA\ region, 
(f) 8333.5--8336~\AA\ region, 
(g) 8726--8730~\AA\ region, 
(h) 8001--8006~\AA\ region, 
(i) 7770--7777~\AA\ region,
(j) 6157--6164~\AA\ region,
(k) 6756--6758.1~\AA\ region, and
(l) 6361--6365~\AA\ region.
Note that the telluric lines in the 8001--8006~\AA\ region, 
8333.5--8336~\AA\ region, and 6361--6365~\AA\ region 
were removed by dividing by the spectra of a rapid rotator.
Likewise, the broad Ca~{\sc i} autoionization feature in the 
6361--6365~\AA\ region was planarized by regarding it as
a pseudo-continuum.
}
\label{fig5}
\end{center}
\end{minipage}
\end{figure}

\subsection{Equivalent width analysis}

As done in T08, we also carried out a differential analysis relative to the Sun 
for the other 19 elements (Al, Si, K, Ca, Sc, Ti, V, Cr, Mn, Co, Ni, Cu, Sr, Y, 
Zr, La, Ce, Pr, and Nd) based on the equivalent widths of usable spectral lines, 
which were measured directly from the OAO spectra by the Gaussian fitting method. 
The procedures of this analysis are detailed in Section~4.1 of Takeda et al. (2005). 
The solar equivalent widths used as the reference for this analysis 
were derived similarly by Gaussian-fitting technique on Kurucz et al.'s (1984) 
solar flux spectrum. The resulting [X/H] values (line-by-line abundances 
relative to the Sun) for $\gamma$~Leo A and B are presented in ``ewanalys\_A.dat'' 
and ``ewanalys\_B.dat'' of the online material, respectively.

It is worth noting that all these abundance results derived from 
equivalent widths are based on the conventional assumption that 
the line opacity is represented by a symmetric Voigt profile.
However, lines of specific groups (e.g., Fe-peak elements of odd atomic number) 
are known to split into a number of sub-components (hyper-fine splitting), 
though the nature of splitting and its significance widely differs 
from case to case. In any event, care should be taken in interpreting 
the results obtained from such lines based on the single-line 
approximation. This effect is separately discussed for the 
representative cases of Sc, V, Mn, Co, and Cu lines in the Appendix.        

\subsection{Final abundances}

Combining what has been described in Sect.~4.2--4.4, the final results of the 
differential abundances relative to the Sun for 26 elements (except for
Li and Be, which are the special cases and thus separately treated in Sect.~5.1.1) 
derived for $\gamma$~Leo A and B are summarized in Table~7, where the differences 
between A and B ($\delta$[X/H]$_{{\rm A}-{\rm B}}$) are also given. The discussions 
presented in Sect.~5.1 and 5.2 will be primarily based on these data.

\setcounter{table}{6}
\setlength{\tabcolsep}{3pt}
\begin{table*}[h]
\scriptsize
\caption{Elemental abundances of both components and their differences.}
\begin{center}
\begin{tabular}{ccccccccccc}\hline\hline
$Z$ & Species & $T_{\rm c}$ & Method & $N_{\rm A}$ & [X/H]$_{\rm A}$ & $\epsilon_{\rm A}$ & 
$N_{\rm B}$ & [X/H]$_{\rm B}$ & $\epsilon_{\rm B}$ & $\delta$[X/H]$_{{\rm A}-{\rm B}}$ \\
(1) & (2) & (3) & (4) & (5) & (6) & (7) & (8) & (9) & (10) & (11) \\
\hline
  6 & C~{\sc i}  &    40 &   fit  &    3 &  $-$0.62 &  0.07 &     3 &  $-$0.64 &  0.04 &   +0.02\\
  7 & N~{\sc i}  &   123 &   fit  &    1 &  $-$0.14 & $\cdots$&     1 &  $-$0.07 & $\cdots$&   $-$0.07\\
  8 & O~{\sc i}  &   180 &   fit  &    1 &  $-$0.33 & $\cdots$&     1 &  $-$0.33 & $\cdots$&   +0.00\\
 11 & Na~{\sc i} &   958 &   fit  &    1 &  $-$0.24 & $\cdots$&     1 &  $-$0.37 & $\cdots$&   +0.13\\
 13 & Al~{\sc i} &  1653 &   eqw  &    2 &  $-$0.21 &  0.02 &     2 &  $-$0.22 & $\cdots$&   +0.01\\
 14 & Si~{\sc i} &  1310 &   eqw  &   33 &  $-$0.21 &  0.02 &    32 &  $-$0.25 &  0.02 &   +0.04\\
 16 & S~{\sc i}  &   664 &   fit  &    1 &  $-$0.36 & $\cdots$&     1 &  $-$0.45 & $\cdots$&   +0.09\\
 19 & K~{\sc i}  &  1006 &   eqw  &    1 &  $-$0.10 & $\cdots$&     1 &  $-$0.07 & $\cdots$&   $-$0.03\\
 20 & Ca~{\sc i} &  1517 &   eqw  &    7 &  $-$0.31 &  0.02 &     6 &  $-$0.30 &  0.01 &   $-$0.01\\
 21 & Sc~{\sc ii}&  1659 &   eqw  &    7 &  $-$0.32 &  0.03 &     8 &  $-$0.26 &  0.04 &   $-$0.06\\
 22 & Ti~{\sc i} &  1582 &   eqw  &   35 &  $-$0.31 &  0.02 &    34 &  $-$0.32 &  0.02 &   +0.01\\
 23 & V~{\sc i}  &  1429 &   eqw  &    8 &  $-$0.32 &  0.02 &     8 &  $-$0.30 &  0.02 &   $-$0.02\\
 24 & Cr~{\sc i} &  1296 &   eqw  &   18 &  $-$0.43 &  0.03 &    18 &  $-$0.39 &  0.02 &   $-$0.04\\
 25 & Mn~{\sc i} &  1158 &   eqw  &    3 &  $-$0.42 &  0.06 &     3 &  $-$0.28 &  0.14 &   $-$0.14\\
 26 & Fe~{\sc i} &  1334 &   eqw  &  191 &  $-$0.41 &  0.03 &   210 &  $-$0.38 &  0.02 &   $-$0.03\\
 27 & Co~{\sc i} &  1352 &   eqw  &    8 &  $-$0.36 &  0.05 &     8 &  $-$0.33 &  0.03 &   $-$0.03\\
 28 & Ni~{\sc i} &  1353 &   eqw  &   41 &  $-$0.41 &  0.02 &    42 &  $-$0.40 &  0.01 &   $-$0.01\\
 29 & Cu~{\sc i} &  1037 &   eqw  &    1 &  $-$0.31 & $\cdots$&     1 &  $-$0.36 & $\cdots$&   +0.05\\
 30 & Zn~{\sc i} &   726 &   fit  &    1 &  $-$0.22 & $\cdots$&     1 &  $-$0.31 & $\cdots$&   +0.09\\
 38 & Sr~{\sc i} &  1464 &   eqw  &    1 &  $-$0.20 & $\cdots$&     1 &  $-$0.23 & $\cdots$&   +0.03\\
 39 & Y~{\sc ii} &  1659 &   eqw  &    3 &  $-$0.40 &  0.07 &     3 &  $-$0.36 &  0.06 &   $-$0.04\\
 40 & Zr~{\sc i} &  1741 &   eqw  &    1 &  $-$0.32 & $\cdots$&     1 &  $-$0.34 & $\cdots$&   +0.02\\
 57 & La~{\sc ii}&  1578 &   eqw  &    1 &  $-$0.09 & $\cdots$&     1 &  $-$0.16 & $\cdots$&   +0.07\\
 58 & Ce~{\sc ii}&  1478 &   eqw  &    4 &  $-$0.26 &  0.01 &     4 &  $-$0.21 &  0.04 &   $-$0.05\\
 59 & Pr~{\sc ii}&  1582 &   eqw  &    2 &  $-$0.32 &  0.11 &     2 &  $-$0.33 &  0.03 &   +0.01\\
 60 & Nd~{\sc ii}&  1602 &   eqw  &    5 &  $-$0.06 &  0.07 &     5 &  $-$0.12 &  0.06 &   +0.06\\
\hline
\end{tabular}
\end{center}
(1) Atomic number. (2) Element species. (3) Condensation temperature (in K) taken from
Table~8 (50\% $T_{\rm c}$) of Lodders (2003). (4) Method for deriving the abundance:
``fit'' $\cdots$ spectrum fitting analysis , ``eqw'' $\cdots$ use of measured equivalent widths. 
(5) Number of available lines (or features) of each species for A.
(6) Final [X/H] value (averaged differential abundance of element X relative to the Sun; in dex) 
for A. (7) Mean error of [X/H] ($\epsilon \equiv \sigma /\sqrt{N}$) for A (in dex).
(8) Number of available lines (or features) for B. (9) Final [X/H] value for B.
(10) Mean error of [X/H] for B.
(11) Differential abundance of A relative to B
 ($\equiv [{\rm X/H}]_{\rm A} - [{\rm X/H}]_{\rm B}$; in dex). 
\end{table*}

\section{Discussion}

\subsection{Light element abundances in context of theoretical predictions}

We first discuss the abundances of light elements, which are expected to have 
suffered more or less changes from the initial composition, because nuclear-processed 
products are dredged-up to the surface by evolution-induced mixing of red giants.  

\subsubsection{Li and Be}

Regarding lithium, our fitting analysis in the Li~{\sc i} 6708 region
resulted in converged solutions at $A_{\rm Li}$ = $-0.81$ (A) and 
$-0.36$ (B). However, these abundances must not be seriously taken 
because they both should be regarded rather as upper limits, 
since the corresponding equivalent widths (1.6 and 0.9~m\AA) are
comparable to or lower than the detection-limit value of a few m\AA\
(cf. Appendix~2 of T17, where it was remarked that $A_{\rm Li} \la 0$ 
is below the reliability limit). Actually, the Li~{\sc i} 6708 line 
feature is too weak to be recognizable by an eye(cf. Fig.~5b) 
What can be said about the surface lithium abundances of $\gamma$~Leo A 
and B is that they have suffered considerable depletion due to an 
efficient envelope mixing in the past (cf. Fig.~19 in T17 for reference).     

As to beryllium, the abundance results are unfortunately less reliable,
because the Be~{\sc ii} 3131 feature is seriously blended with the neighboring
Fe line (owing to the appreciably large macroturbulence in this UV region 
presumably due to its height-increasing nature) and the Fe abundance
had to be fixed (i.e. simultaneous determination with Be was not possible). 
Given this in mind, the derived $A_{\rm Be}$ values of $-0.23$ (A) and 
$0.08$ (B) correspond to [Be/Fe] = $-1.24$ (A) and $-0.96$ (B) 
(assuming $A_{\rm Be\odot} = 1.42$ as in T14), which suggests that
the Be deficiency is more or less compatible with the theoretical
prediction expecting [Be/Fe]~$\sim -1$ to $-2$ (see the orange line 
corresponding to $M = 1.5 M_{\odot}$ in Fig.~6 of T14).  

\subsubsection{C, N, O, and Na}

The key elements, the abundances of which may be affected by the dredge-up 
of H-burning products are C, N, O (CNO-cycle), and Na (NeNa-cycle).
The relative abundance ratios of [C/Fe], [N/Fe], [O/Fe], and [Na/Fe] 
for $\gamma$~Leo A/B are $-0.21/-0.26$, $+0.27/+0.31$, $+0.08/+0.05$, 
and $+0.17/+0.01$, respectively, The characteristics (almost similar 
for both A and B) are: (i) [C/Fe] is subsolar by $\sim$~0.2--0.3~dex, 
(ii) [N/Fe] is supersolar by $\sim 0.3$~dex, (iii) [O/Fe] is slightly 
supersolar by $\la 0.1$~dex, and [Na/Fe] is by $\la 0.2$~dex supersolar 
(A) or almost solar (B). Likewise, an appreciably low $^{12}$C/$^{13}$C
ratio around $\sim 10$ was obtained for both stars.

These trends are more or less (at least qualitatively) consistent with 
the theoretical expectations (cf. Fig.~11 in T15 for C, O, and Na; 
Fig.~12 in T19 for N and $^{12}$C/$^{13}$C). The slightly positive [O/Fe] 
(despite that O may suffer a very slight deficiency by $\la$ a few 
hundredths dex; cf. Fig.~11b) is attributed to the chemical evolution 
effect for a mildly lower metallicity of [Fe/H] $\sim -0.4$.
Accordingly, we may state that the abundance characteristics
of these light elements are reasonably explained by the canonical
theory of stellar evolution.

\subsubsection{Comparison with previous work}

Lambert \& Ries's (1981) pioneering study of CNO abundances for 32 G--K 
giants (based on [O~{\sc i}] lines and C$_{2}$ as well as CN molecular lines) 
included both $\gamma$~Leo~A/$\gamma$~Leo~B, for which they obtained
$-0.26/-0.31$ for [C/Fe], $+0.42/+0.25$ for [N/Fe],
$+0.17/+0.12$ for [O/Fe], and $6.5/9$ for $^{12}$C/$^{13}$C.
These values (along with the metallicities of [Fe/H] =~$-0.35/-0.37$) 
are mostly consistent with our results, despite that the adopted lines 
are different. 

However, a few previous studies done in 1990s reported apparently 
contradicting results regarding the C and N abundances. 
That is, a slightly subsolar [N/Fe] of $\sim -0.1$ and a somewhat supersolar
[C/Fe] of $\sim +0.2$ were derived for $\gamma$~Leo A by Shavrina et al. (1996a)
and Shavrina et al. (1996b) based on NH bands (around $\sim 3360$~\AA) 
and CH bands (at 4230--4270~\AA), which are just the opposite trend to
what was obtained by Lambert \& Ries (1981) and in this paper. 
Then, an independent study (again based on the NH bands around $\sim 3360$~\AA)
was soon after carried out by Yakovina \& Pavlenko (1998), who concluded 
[N/Fe] = $0.0 (\pm 0.1)$ for $\gamma$~Leo~A (slightly higher by 0.1~dex than Shavrina 
et al.'s result). In any event, the C and N abundances of $\gamma$~Leo~A derived by
these groups from the spectrum-synthesis analysis of CH and NH bands in the blue--UV 
region are in conflict with the standard stellar evolution theory, which predicts 
a C-deficiency as well as an N-enrichment by a few tenths dex.
Although any definite argument can not be made, their conclusions seem to be questionable. 
Since absolute CNO abundances determined from the molecular bands of hydride 
molecules (CH, NH, and OH) in short wavelength regions tend to suffer systematic 
errors, these errors for the target red giant and the Sun may not have been  
well cancelled out in the resulting differential abundances ([C/H] and [N/H]). 
Anyway, this problem is worth further investigation.  

\setcounter{figure}{5}
\begin{figure}[H]
\begin{minipage}{80mm}
\begin{center}
\includegraphics[width=6.0cm]{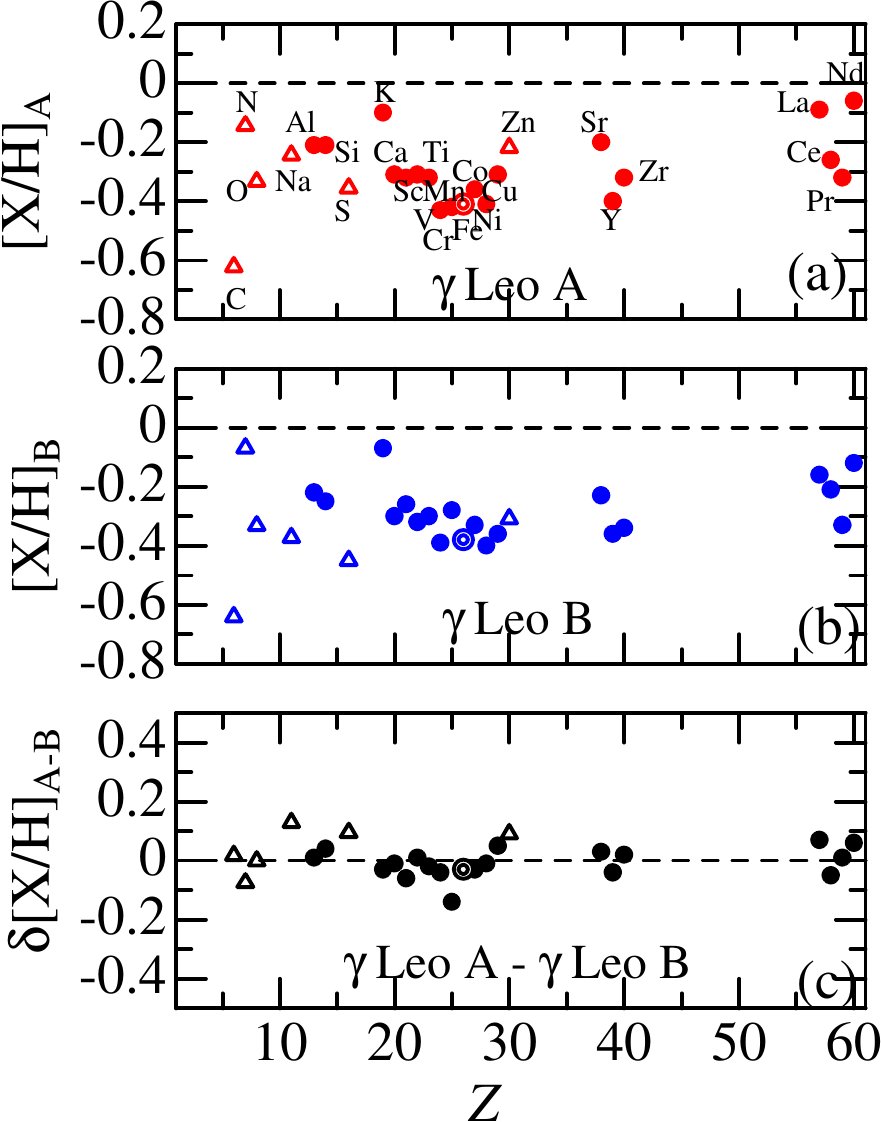}
\caption{
In panels (a) and (b) are plotted the [X/H] values (logarithmic abundance 
of element X relative to the Sun) against the atomic number ($Z$) 
for $\gamma$~Leo A and B, respectively. The differences between A and B 
($\delta$[X/H]$_{{\rm A}-{\rm B}} \equiv$~[X/H]$_{\rm A}-$[X/H]$_{\rm B}$) 
are also shown in the bottom panel (c). 
The results derived from the spectrum synthesis analysis are shown by triangles 
while those from the equivalent width analysis are by circles (the data points 
for Fe are depicted by double circles). Note that the results for Li and 
Be are not shown here.
}
\label{fig6}
\end{center}
\end{minipage}
\end{figure}

\subsection{Any abundance difference between A and B?}

Searching for characteristic chemical signatures in stars specific to their 
planet-harboring nature (such as metal-rich tendency, abundance differences 
between volatile and refractory species; etc.) is important, since they may 
serve as a key to clarifying the physical mechanism of planet formation.
A number of spectroscopic investigations on the chemical abundances of 
planet-host stars (along with the comparison sample of non-planet-host stars) 
have been done toward this aim over the past quarter century. 
Although solar-type stars (FGK-type dwarfs) have been primarily targeted in 
most of these studies, several recent papers have focused also on evolved red 
giants (retired A-type stars) reflecting the increasing number of planets 
discovered around giants; e.g., Maldonado, Villaver, \& Eiroa (2013), 
da Silva et al. (2015), and Jofr\'{e} et al. (2015), and Maldonado \& Villaver (2016). 
While some possible characteristic trends specific to planet-host giants 
are reported compared to normal non-planet-host giants (e.g., possible metal-rich 
tendency in the higher-mass regime, small difference in [X/Fe] for some elements), 
they are generally subtle and indistinct. Moreover, it is not easy to 
discern the effect due to the existence of planets from the systematic 
problem caused by the difference between the samples.   

In contrast, our approach is more straightforward and simple:
To examine whether any difference exists in the elemental abundances 
between $\gamma$~Leo~A (with planet) and $\gamma$~Leo~B (without planet),
both of which are considered to have formed from the gas with the same composition.   
Based on the results summarized in Table~7, the final abundances relative 
to the Sun for 26 elements from C to Nd ([X/H]$_{\rm A}$, [X/H]$_{\rm B}$, 
and their differences $\delta$[X/H]$_{{\rm A}-{\rm B}}$) are plotted against 
the atomic number in Fig.~6a--c.
As seen from Table~7 and Fig.~6c, $|\delta$[X/H]$_{{\rm A}-{\rm B}}| \la 0.1$~dex 
holds for almost all of the studied elements.

The abundance errors due to ambiguities in atmospheric parameters 
($T_{\rm eff}$, $\log g$, and $v_{\rm t}$) can be estimated for C, N, O,
Na, S, and Zn (fitting-based abundances) by consulting
Table~3 in T15, Table~3 in T16, and Table~2 in T19.
Adopting $\pm 20$~K, $\pm 0.1$~dex, and $\pm 0.1$~km~s$^{-1}$ for the
uncertainties in $T_{\rm eff}$, $\log g$, and $v_{\rm t}$ (cf. Sect. 3.1),
we obtained the corresponding abundance errors (root-sum-square of 
three error components) as $\pm 0.04$~dex for 
C~{\sc i} 5380 (though this line was eventually discarded, it is typical 
for high-excitation C~{\sc i} lines such as C~{\sc i} 5052 and C~{\sc i} 8335),
$\pm 0.06$~dex for [C~{\sc i}] 8727,
$\pm 0.04$~dex for N (from CN~8003), 
$\pm 0.06$~dex for O~{\sc i} 7774, 
$\pm 0.03$~dex for Na~{\sc i} 6161,
$\pm 0.06$~dex for S~{\sc i} 6757, and
$\pm 0.04$~dex for Zn~{\sc i} 6362.
Meanwhile, regarding the abundances for many other elements derived from directly 
measured equivalent widths, their mean errors ($\epsilon$) given in Table~7 are 
mostly within $\la$~0.06--0.07~dex (exceptionally as large as $\sim 0.1$~dex 
for a few cases).
We may thus regard that the errors involved in both [X/H]$_{\rm A}$ and
[X/H]$_{\rm B}$ values are $\la 0.1$~dex, which means that the
 abundances of $\gamma$~Leo~A (with planet) and B (without planet) are 
practically the same (i.e., the $|\delta$[X/H]$_{{\rm A}-{\rm B}}|$ values 
of $\la 0.1$~dex are within the uncertain range). 
This conclusion may imply that the existence of a planet around A does 
not have any appreciable impact on the current surface chemical composition.

Our result is in agreement with the available previous work 
(cf. Table 1). Lambert \& Ries's (1981) CNO analysis for A and B 
resulted in almost the same similarity (cf. Sect.~5.1.3).
Likewise, the same consequence can be drawn from the abundances of 
$\gamma$~Leo A and B derived by McWilliam (1990): Table~13 in his paper 
shows that the values of $\langle \delta A_{\rm{A}-\rm{B}} \rangle$ 
(mean differential abundances between A and B averaged over available lines) 
are +0.14, 0.00, +0.02, $-0.07$, +0.05, $-0.04$, $+0.04$, +0.06, $-0.06$, 
+0.06, +0.03, and +0.05~dex for Si~{\sc i}, Ca~{\sc i}, Sc~{\sc ii}, Ti~{\sc i}, 
Ti~{\sc ii}, V~{\sc i}, Co~{\sc i}, Ni~{\sc i}, Y~{\sc ii}, La~{\sc ii}, 
Nd~{\sc ii}, and Eu~{\sc ii}, respectively.   

Finally, the [X/Fe] vs. $T_{\rm c}$ plots for A and B are also displayed in Fig.~7. 
Although this diagram is known to be useful in searching for any chemical 
signature of proto-planetary material accreted at the time of planet formation 
(i.e., abundance difference between volatile elements of low $T_{c}$
and refractory elements of high $T_{\rm c}$), its interpretation is difficult in the 
present case of red giants, because CNO abundances (representative low-$T_{\rm c}$ 
elements) tend to suffer changes due to an evolution-induced mixing. 
Anyway, no apparent $T_{\rm c}$-dependent trend is observed in [X/Fe] 
for both A and B (Fig.~7a and Fig.~7b)..

\setcounter{figure}{6}
\begin{figure}[H]
\begin{minipage}{80mm}
\begin{center}
\includegraphics[width=6.0cm]{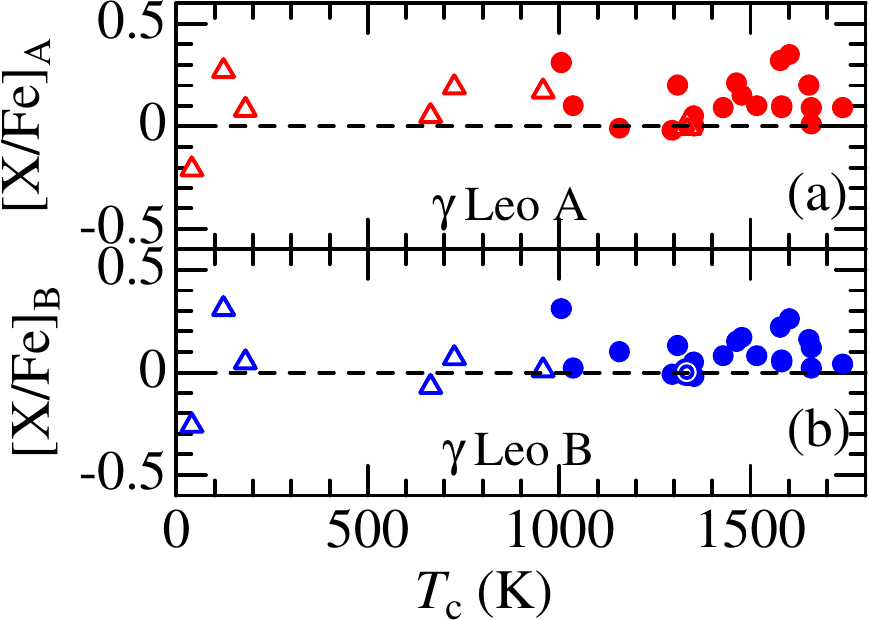}
\caption{
The [X/Fe] values ($\equiv [{\rm X/H}]- [{\rm Fe/H}]$; logarithmic X-to-Fe
abundance ratios relative to the Sun) are plotted against $T_{\rm c}$ 
(condensation temperature).
Panels (a) and (b) are for $\gamma$~Leo A and B, respectively.
Otherwise, the same as in Fig.~6.
}
\label{fig7}
\end{center}
\end{minipage}
\end{figure}

\subsection{Mass-related problems}

As summarized in Table~2, the mass values were derived in Sect.~3.2 as 
1.66~$M_{\odot}$ (A) and 1.55~$M_{\odot}$ (B) from the positions on the HR 
diagram in comparison with theoretical evolutionary tracks. 
Since the corresponding $\log g_{MR}$ (1.80 and 2.56; evaluated from 
$M$ and $R$) are consistent with the spectroscopic $\log g$ 
(1.89 and 2.53) and the resulting $age$s for A and B are in agreement with 
each other at $\sim 2$~Gyr within the error bars (cf. Fig.~3), 
we may regard that these masses are reasonable.

As mentioned in Sect~1,  previously determined mass values of $\gamma$~Leo~A 
are considerably diversified from $\sim 1.2 M_{\odot}$ to $\sim 1.8 M_{\odot}$
(cf. Table~1). Especially, since the value of 1.23~$M_{\odot}$ adopted by 
Han et al. (2010) is presumably too low, the mass of the planet discovered 
by them ($m_{\rm p} \sin i_{\rm p} = 8.78 M_{\rm Jupiter}$) should be 
revised upward by a factor of $(1.66/1.23)^{2/3}$ as 
$m_{\rm p} \sin i_{\rm p} \simeq 10.7 M_{\rm Jupiter}$. 
This value has got closer to the critical demarcation mass ($13 M_{\rm Jupiter}$)  
between planet and brown dwarf, increasing a possibility that this substellar 
object orbiting around $\gamma$~Leo A may rather fall in the category of 
brown dwarf (depending on the inclination angle $i_{\rm p}$). 

As already remarked in Sect.~1, the orbital elements of $\gamma$~Leo system are still 
subject to large uncertainties because of its very long period over several centuries.
For example, according to Burnham's Celestial Handbook (Burnham 1978),
determinations of its orbital period in old literature are considerably
diversified from $\sim 400$~yr to $\sim 700$~yr. 
The Sixth Catalog of Orbits of Visual Binary Stars\footnote{
Available at https://crf.usno.navy.mil/wds-orb6.} (WDS-ORB6; cf. Hartkopf et al. 2001)
gives $P = 554 (\pm 27)$~yr (period), $a = 3.10 (\pm 0.10)$~arcsec (semimajor axis),
$i = 41 (\pm 5)^{\circ}$ (inclination angle), and $e = 0.93 (\pm 0.02)$ (ellipticity) for 
$\gamma$~Leo, though a low grade `4 (preliminary)' is assigned to these data.
Then, revised elements were published by Mason et al. (2006; cf. Table~7 therein),
as $P = 510.3$~yr, $a = 4.24$~arcsec, $i = 76.0^{\circ}$, and $e = 0.845$ 
(again with grade `4'). 
Following Kepler's third law, the sum of the masses in a binary system
($M_{\rm A} + M_{\rm B}$ in unit of $M_{\odot}$) is expressed 
in terms $a$ (in arcsec), $\pi$ (parallax; in arcsec), and 
$P$ (in yr) as $M_{\rm A} + M_{\rm B} = a^{3} \pi^{-3} P^{-2}$. 
However, this relation (with $\pi = 25.07 \times 10^{-3}$ arcsec) 
yields $M_{\rm A} + M_{\rm B} = 6.2~M_{\odot}$ 
(in case of WDS-ORB6 elements) or $18.6 M_{\odot}$ (in case of Mason et al.'s elements), 
which seriously disagree with our result of $M_{\rm A} + M_{\rm B} = 3.21 M_{\odot}$ 
obtained in this study.
This discrepancy manifestly suggests that the published orbital elements 
of the $\gamma$~Leo system (even the latest ones) should be viewed with caution. 
Further long-running observations (at least over the next hundreds of years) would 
be required before this problem could be settled.
 
\section{Summary and conclusion}

$\gamma$~Leo is a binary system comprising two similar red giants of 
A (with planet) and B (without planet).
It is worthwhile to examine if any difference exists between 
the surface abundances between these two, which may provide some information 
on a possible impact of planet formation upon the chemistry of the host star.

Yet, spectroscopic studies intending to clarify the chemical properties for 
both A and B seem to have been barely conducted so far. This motivated
the author to newly determine the abundances of many elements for both components 
based on the high-dispersion spectra covering wide wavelength ranges.

Besides, the stellar parameters and related characteristics (e.g., stellar 
activity, kinematic information, etc.) were also studied because they are 
not necessarily well established, where particular attention was paid to 
clarifying the stellar mass (for which published results are diversified).

In addition, the nature of internal mixing in these evolved stars could be
checked based on the light element abundances, because conflicting 
arguments were made by different authors regarding the surface abundances 
of C and N in $\gamma$~Leo~A (i.e., whether or not they are well explained by 
the standard theory for the evolution-induced dredge-up of H-burning products).

The atmospheric parameters ($T_{\rm eff}$, $\log g$, and $v_{\rm t}$) were 
spectroscopically determined from Fe~{\sc i} and Fe~{\sc ii} lines. The 
resulting Fe abundances are [Fe/H] = $-0.41$ (A) and $-0.38$ (B); i.e.,
almost the same at a mildly subsolar metallicity.
The kinematic parameters suggest that this system belongs to the 
thin-disk population.

The masses were derived from the positions on the HR diagram in comparison with
theoretical evolutionary tracks as 1.66~$M_{\odot}$ (A) and 1.55~$M_{\odot}$ (B).
Both A and B are likely to be in the stage of red clump giants after He-ignition. 
According to the newly determined $M_{\rm A}$, the mass of the planet around A 
was also revised as $m_{\rm p} \sin i_{\rm p} \simeq 10.7 M_{\rm Jupiter}$ 
(increased by $\sim 20\%$ from the original value reported by Han et al.). 

The chromospheric activity was estimated from the core emission strength of 
the Ca~{\sc ii} 3934 line, and the projected rotational velocity
($v_{\rm e}\sin i$) was determined from the spectrum-fitting analysis.
It turned out that both $\gamma$~Leo A and B are typical red giant stars of 
low activity and low rotational velocities.

Chemical abundances of especially important elements (Li, Be, C, N, O, Na, S, 
and Zn;  either being affected by evolution-induced mixing or volatile species)
were determined by the spectrum-fitting technique. Meanwhile, those of the remaining elements
(Al, Si, K, Ca, Sc, Ti, V, Cr, Mn, Co, Ni, Cu, Sr, Y, Zr, La, Ce, Pr, and Nd; 
mostly refractory species) were derived by the conventional method using 
the directly measured equivalent widths.

Regarding the mixing-affected light elements, 
although much can not be said about Li (very depleted; only upper limit)
and Be (considerably deficient by $\sim -1$~dex, though less reliable) ,
a moderate deficiency of C, 
a mild enrichment of N, and a slight overabundance of Na, and
a low $^{12}$C/$^{13}$C ratio were obtained for both A and B, which are 
quite consistent with the trend expected from the canonical theory of 
stellar evolution. Likewise, the slightly positive [O/Fe] is reasonable for 
these somewhat metal-poor stars (galactic chemical evolution effect).

The chemical abundances of A and B turned out to be practically the same 
within $\la 0.1$~dex for almost all elements, which implies that the surface 
chemistry is not appreciably affected by the existence of a planet in A.
Likewise, any meaningful $T_{\rm c}$-dependent trend (or difference
between volatile and refractory species) in [X/Fe] was not observed. 

Accordingly, what can be concluded from this investigation is as follows. 
\begin{itemize}
\item
Based on the results summarized above, we may state that the visual 
binary system $\gamma$~Leo A+B is not so spectroscopically unusual 
as suspected initially when this investigation was motivated. 
\item
The fact that no clear abundance difference was detected between A (with planet) 
and B (without planet) suggests that hosting a planet does not have an 
appreciable impact on the surface abundances of red giants, though this 
is an argument specific to $\gamma$~Leo and may not simply be generalized.
\item
The abundances of light elements are well consistent with those predicted
from the canonical mixing theory of stellar evolution, which means that both 
$\gamma$~Leo A and B are ordinary red giants (presumably of red clump)
like many others. 
\end{itemize}

\section*{Acknowledgments}

This research is in part based on data obtained by the Subaru Telescope, 
operated by the National Astronomical Observatory of Japan.
This investigation has made use of the SIMBAD database, operated by CDS, 
Strasbourg, France, and the VALD database operated at Uppsala University,
the Institute of Astronomy RAS in Moscow, and the University of Vienna.

\newpage

\section*{Data availability}

The large data (equivalent widths, abundances, atomic data) of 
the spectroscopic analysis are given in the electronic data files 
of the supplementary material.
The raw data for the spectra of $\gamma$~Leo A and B used in 
this investigation are in the public domain and available at 
https://smoka.nao.ac.jp/index.jsp (SMOKA Science Archive site).

\section*{Supplementary information}

The following online data are available as supplementary 
materials accompanied with this article. 
\begin{itemize}
\item
{\bf readme.txt} 
\item
{\bf feabunds.dat} 
\item
{\bf ewanalys\_A.dat} 
\item
{\bf ewanalys\_B.dat} 
\end{itemize}

\section*{Statements and declarations}

\subsection*{Funding}
The author declares that no funds, grants, or other support were received 
during the preparation of this manuscript.

\subsection*{Competing interests}
The author has no relevant financial or non-financial interests to disclose.

\subsection*{Author contributions}
This investigation has been conducted solely by the author.

\section*{Appendix: Impact of hyperfine splitting on abundance determination} 

In the determination of the abundances for heavier elements based on the equivalent 
widths described in Sect.~4.4, the conventional single-line treatment was adopted where 
the line opacity is represented by a symmetric Voigt function. While this assumption 
is valid for most cases, lines of some elements (especially odd-$Z$ elements around
$Z \sim$~20--30) are known to intricately split into sub-components, which is caused
by nucleus--electron coupling of the angular momentum.
Since this effect (hyper-file splitting; abbreviated as ``hfs'') acts
as an extra broadening of the line opacity (while the total integrated opacity
being kept unchanged) like the case of microturbulence, the equivalent width 
(of more or less saturated line) is increased by this splitting effect compared 
to the non-split case. As a result, the abundance derived from such a hfs-split 
line based on the usual single-component assumption tends to be overestimated 
unless the line is very weak.

The extents of overestimation caused by applying the single-component approximation 
were examined for the representative hfs lines of Sc~{\sc ii} ($Z=21$),
V~{\sc i} ($Z=23$), Mn~{\sc i} ($Z=25$), Co~{\sc i} ($Z=27$), Cu~{\sc i} ($Z=29$).
Out of the 29 lines for these 5 species analyzed in Sect.~4.4, 
22 lines were selected for this test (cf. Table~8), for which the relevant 
hfs data (wavelengths and relative strengths of subcomponents) are available 
in the ``gfhyperall.dat'' file downloaded from the Dr. R. L. Kurucz's 
web site.\footnote{http://kurucz.harvard.edu/linelists/gfhyperall/}

By using the WIDTH9 program which was so modified as to enable incorporating 
the line-splitting effect by the spectrum-synthesis technique, two kinds of 
abundances [$A$(with hfs) and $A$(without hfs)] were obtained for given 
equivalent widths ($W_{\lambda}$), and the corresponding hfs corrections 
$\delta A [\equiv A$(with hfs)$- A$(without hfs)] 
were computed for $\gamma$~Leo A and B along with the Sun.
Similarly, its effect on [X/H]$_{*}$ (abundance of element X relative to the Sun) 
could be evaluated as $\delta$[X/H]$_{*} \equiv \delta A_{*} - \delta A_{\odot}$.
The results are summarized in Table~8, and in Fig.~8 are plotted these 
$\delta A$ as well as $\delta$[X/H] values against $W_{\lambda}$. 
An inspection of Fig.~8 reveals the following characteristics.
\begin{itemize}
\item
The hfs corrections on the abundances ($\delta A$) are always negative, 
and $|\delta A|$ tends to be larger for stronger lines, as expected.
However, their extents are considerably diversified from case to case depending on 
the detailed nature of line splitting. For example, even in the same strong-line 
regime ($W_{\lambda} \sim 100$~m\AA) for the case of $\gamma$~Leo~A (Fig.~8a),  
$|\delta A|$ is almost negligible for Sc~{\sc ii}~5526.790  
while significantly large ($\sim 0.3$~dex) for V~{\sc i} 5670.853.    
\item
Roughly speaking, the inequality relation 
$|\delta A_{\odot}| < |\delta A_{\rm B}| < |\delta A_{\rm A}|$ 
holds for the relative importance of hfs corrections between 
$\gamma$~Leo A and B and the Sun (Fig.~8a and Fig.~8c). 
Therefore, as to $|\delta$[X/H]$|$ (generally smaller than $|\delta A|$
due to the subtraction of solar correction), $|\delta[{\rm X}/{\rm H}]_{\rm B}|$ 
is distinctly smaller than $|\delta[{\rm X}/{\rm H}]_{\rm A}|$. 
Actually, while $|\delta$[X/H]$_{\rm A}|$ for $\gamma$~Leo~A are still appreciable 
and significant (around $\sim 0.1$~dex on the average, though extending up to 
$\sim 0.3$~dex for some lines; cf. Fig.~8b),  $|\delta$[X/H]$_{\rm B}|$ for 
$\gamma$~Leo~B is insignificant (confined within $\lesssim 0.05$~dex; cf. Fig.~8d).
\item
As seen from the different extent of hfs correction between the three cases 
(Sun $<$ $\gamma$~Leo~B $<$ $\gamma$~Leo~A), $T_{\rm eff}$ is presumably 
the most critical factor determining the significance of hfs, because it 
affects (i) the equivalent width of a line and (ii) the thermal width 
of line opacity, both affecting the degree of saturation. Accordingly, we may 
generally state that the impact of hfs on abundance determinations becomes 
more significant as $T_{\rm eff}$ is lowered.  
\end{itemize}

Then, how much hfs correction should be applied to the relevant abundance 
results of Sc, V, Mn, Co, ad Cu obtained in Sect.~4.4 (which were derived from 
equivalent widths based on the single-line assumption neglecting hfs)?
As long as the lines presented in Table~8 are concerned, while the corrections
($\delta$[X/H]$_{\rm B}$) for $\gamma$~Leo~B are apparently insignificant 
(only a few hundredths dex in any case), appreciable downward corrections ranging 
from $\sim 0.0$ to $\sim 0.3$~dex (considerably differing from line to line) 
are expected for $\gamma$~Leo~A as seen from the values of $\delta$[X/H]$_{\rm A}$. 
Fortunately, even in the latter case of $\gamma$~Leo~A, since the lines of large 
corrections (by $\sim$~0.2--0.3~dex) belong to the species (i.e., V or Co) using 
a sufficient number of lines (8--9), their impact tends to be mitigated 
after averaging. By applying the $\delta$[X/H] corrections of 22 lines (Table~8)  
to the [X/H] values obtained in Sect.~4.4 (cf. ``ewanalys\_A.dat'' and 
``ewanalys\_B.dat'' in the online material), new mean $\langle$[X/H]$\rangle$ 
values (with hfs included) were calculated (where the same [X/H] data were used 
unchanged for the 7 lines for which hfs data were unavailable).

The resulting differences (in dex) of 
$\langle$[X/H]$\rangle$(with hfs)$-$$\langle$[X/H]$\rangle$(without hfs) are
$-0.02|-0.02$ (Sc), $-0.07|0.00$ (V), $0.00|0.00$ (Mn), $-0.07|-0.02$ (Co),
$-0.01|-0.01$ (Cu), for $\gamma$~Leo A$|$B, respectively.
Accordingly, the hfs corrections on the final $\langle$[X/H]$\rangle$ results
of Sc, V, Mn, Co, and Cu derived in Sect.~4.4 are only slight reductions by 
$\lesssim 0.1$~dex and thus not significant.

\newpage

\setcounter{table}{7}
\begin{table*}[h]
\scriptsize
\caption{Hyperfine-splitting effect on the abundances of Sc, V, Mn, Co, and Cu.}
\begin{center}
\begin{tabular}{crrrrrrrr}\hline
Line & $W_{\lambda,\odot}$ & $\delta A_{\odot}$ & 
$W_{\lambda,{\rm A}}$ & $\delta A_{\rm A}$ & $\delta$[X/H]$_{\rm A}$ &
$W_{\lambda,{\rm B}}$ & $\delta A_{\rm B}$ & $\delta$[X/H]$_{\rm B}$ \\
(1) & (2) & (3) & (4) & (5) & (6) & (7) & (8) & (9) \\
\hline
Sc~{\sc ii} 5318.349  &   14.0 &$-$0.001 &    43.4& $-$0.004& $-$0.003  &   33.3& $-$0.003& $-$0.002\\
Sc~{\sc ii} 5526.790  &   75.6 &$-$0.003 &   119.9& $-$0.001& +0.002  &  110.8& $-$0.003&  0.000\\
Sc~{\sc ii} 5552.224  &    5.1 &$-$0.001 &    22.8& $-$0.008& $-$0.007  &   16.9& $-$0.005& $-$0.004\\
Sc~{\sc ii} 5667.149  &   32.4 &$-$0.026 &    76.7& $-$0.079& $-$0.053  &   63.9& $-$0.062& $-$0.036\\
Sc~{\sc ii} 5669.042  &   35.3 &$-$0.002 &    86.6& $-$0.009& $-$0.007  &   76.9& $-$0.008& $-$0.006\\
Sc~{\sc ii} 5684.202  &   37.1 &$-$0.028 &    86.4& $-$0.085& $-$0.057  &   74.6& $-$0.071& $-$0.043\\
Sc~{\sc ii} 6245.637  &   35.3 &$-$0.027 &    83.9& $-$0.079& $-$0.052  &   71.6& $-$0.065& $-$0.038\\
V~{\sc i} 5657.435  &    6.5 &$-$0.001 &    67.5& $-$0.048& $-$0.047  &   20.6& $-$0.008& $-$0.007\\
V~{\sc i} 5668.361  &    6.6 &$-$0.005 &    67.0& $-$0.096& $-$0.091  &   21.7& $-$0.018& $-$0.013\\
V~{\sc i} 5670.853  &   19.7 &$-$0.023 &   118.3& $-$0.290& $-$0.267  &   52.4& $-$0.071& $-$0.048\\
V~{\sc i} 6224.529  &    6.4 &$-$0.015 &    91.4& $-$0.343& $-$0.328  &   25.9& $-$0.057& $-$0.042\\
V~{\sc i} 6266.307  &    3.3 &$-$0.003 &    59.7& $-$0.094& $-$0.091  &   24.6& $-$0.031& $-$0.028\\
V~{\sc i} 6326.840  &    2.3 &$-$0.004 &    26.8& $-$0.092& $-$0.088  &    7.8& $-$0.028& $-$0.024\\
V~{\sc i} 6531.415  &    6.9 &$-$0.007 &    69.5& $-$0.100& $-$0.093  &   19.9& $-$0.016& $-$0.009\\
V~{\sc i} 7338.942  &    2.6 &$-$0.010 &    20.0& $-$0.065& $-$0.055  &    6.9& $-$0.020& $-$0.010\\
Mn~{\sc i} 6440.971  &    6.4 &$-$0.003 &    15.9& $-$0.008& $-$0.005  &    7.4& $-$0.003&  0.000\\
Co~{\sc i} 5280.629  &   19.7 &$-$0.090 &    53.0& $-$0.283& $-$0.193  &   32.7& $-$0.139& $-$0.049\\
Co~{\sc i} 5301.039  &   20.1 &$-$0.019 &    93.3& $-$0.166& $-$0.147  &   55.3& $-$0.064& $-$0.045\\
Co~{\sc i} 5342.695  &   31.8 &$-$0.146 &    59.1& $-$0.303& $-$0.157  &   43.7& $-$0.186& $-$0.040\\
Co~{\sc i} 5454.572  &   13.4 &$-$0.042 &    27.7& $-$0.081& $-$0.039  &   19.3& $-$0.052& $-$0.010\\
Co~{\sc i} 6477.853  &    4.2 &$-$0.013 &    16.0& $-$0.042& $-$0.029  &    8.7& $-$0.021& $-$0.008\\
Cu~{\sc i} 5218.197  &   52.7 &$-$0.006 &    80.4& $-$0.017& $-$0.011  &   65.3& $-$0.014& $-$0.008\\
\hline
\end{tabular}
\end{center}
(1) Line designation. (2) Solar equivalent width (in m\AA). (3) Correction of the hyperfine-splitting 
effect to the solar abundance (in dex), which is defined as $\delta A \equiv A$(with hfs)$-A$(without hfs).
(4) Equivalent width of $\gamma$~Leo~A. (5) Hfs correction to the abundance of $\gamma$~Leo~A.
(6) Hfs correction to the [X/H] of $\gamma$~Leo~A ($\equiv \delta A_{\rm A} - \delta A_{\odot}$).
(7) Equivalent width of $\gamma$~Leo~B. (8) Hfs correction to the abundance of $\gamma$~Leo~B.
(9) Hfs correction to the [X/H] of $\gamma$~Leo~B.
\end{table*}

\setcounter{figure}{7}
\begin{figure}[H]
\begin{minipage}{80mm}
\begin{center}
\includegraphics[width=7.0cm]{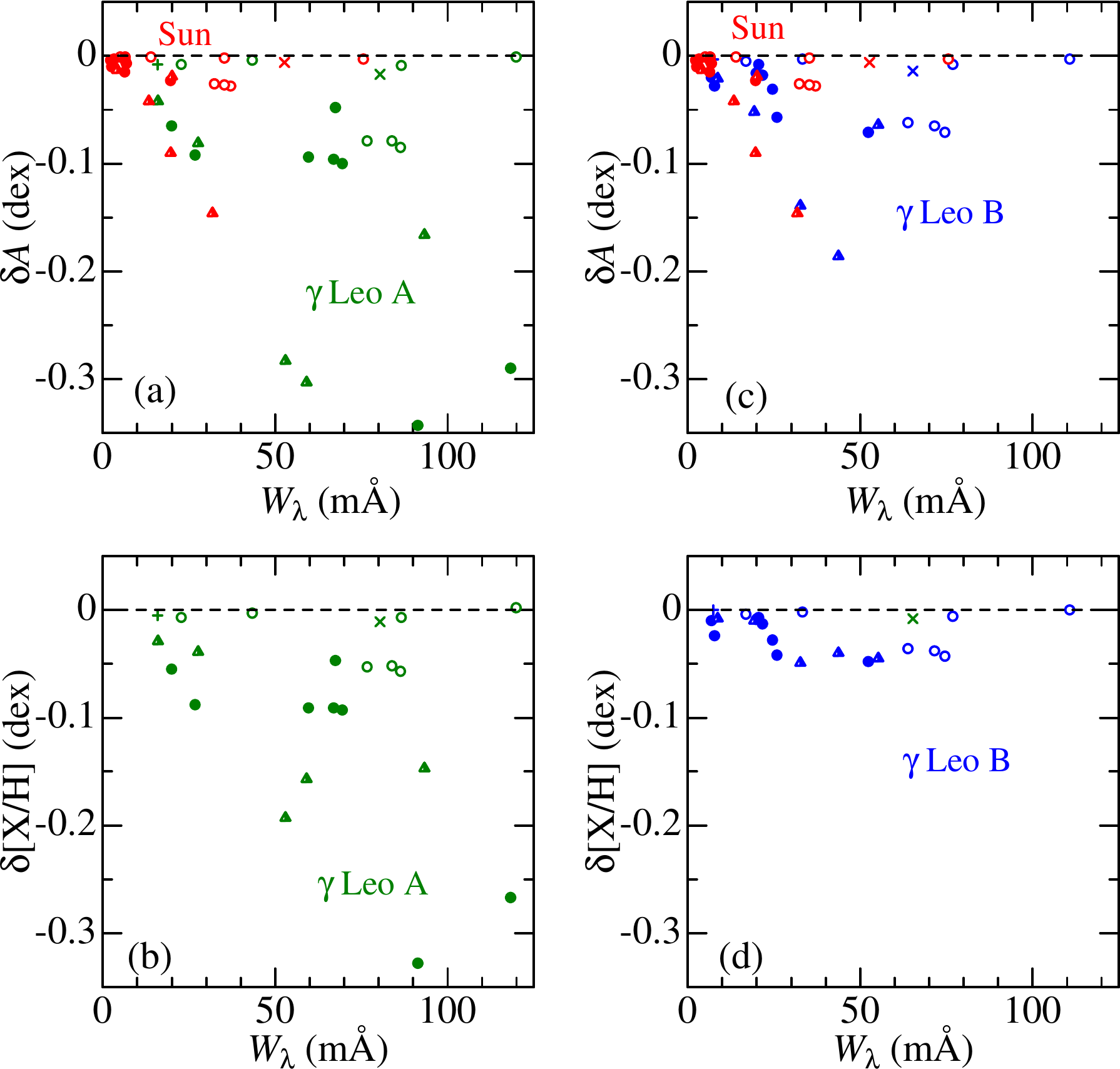}
\caption{
Corrections to the absolute ($A$) or relative ([X/H]) abundances due to the 
hyperfine-splitting effect (cf. Table~8) are plotted against the equivalent widths, 
where the open circles, filled circles, Greek crosses (+), half-filled triangles, 
and St. Andrew's crosses ($\times$) correspond to Sc, V, Mn, Co, and Cu, respectively.
(a) Hfs corrections to $A$ for $\gamma$~Leo~A (green) and the Sun (red). 
(b) Hfs corrections to [X/H] for $\gamma$~Leo~A.
(c) Hfs corrections to $A$ for $\gamma$~Leo~B (blue) and the Sun (red). 
(d) Hfs corrections to [X/H] for $\gamma$~Leo~B.
}
\label{fig8}
\end{center}
\end{minipage}
\end{figure}

\end{document}